\documentclass[lettersize,journal]{IEEEtran}
\usepackage{cite}
\usepackage{amsmath,amssymb,amsfonts}
\usepackage{algorithm}
\usepackage{algorithmic}
\usepackage{array}
\usepackage[caption=false,font=normalsize,labelfont=sf,textfont=sf]{subfig}
\usepackage{textcomp}
\usepackage{stfloats}
\usepackage{url}
\usepackage{verbatim}
\usepackage{graphicx}
\def\BibTeX{{\rm B\kern-.05em{\sc i\kern-.025em b}\kern-.08em
    T\kern-.1667em\lower.7ex\hbox{E}\kern-.125emX}}
\usepackage{balance}
\begin{document}

\title{Federated Multi-View Synthesizing for Metaverse}
\author{Yiyu Guo,~\IEEEmembership{Graduate Student Member,~IEEE}, Zhijin Qin,~\IEEEmembership{Senior Member,~IEEE}, Xiaoming Tao,~\IEEEmembership{Senior Member,~IEEE}, and Geoffrey Ye Li,~\IEEEmembership{Fellow,~IEEE}

\thanks{Manuscript received 14 March 2023; revised 24 July 2023 and 29 September 2023. This work was supported in part by the National Natural Science Foundation of China (NSFC) under Grant 62293484 and Grant 61925105. Part of the work was presented at the 2022 IEEE 96th Vehicular Technology Conference\cite{10012859}. (Corresponding author: Zhijin Qin.)}
\thanks{Yiyu Guo is with the School of Electronic Engineering and Computer Science, Queen Mary University of London, London E1 4NS, U.K. (email: yiyu.guo@qmul.ac.uk)}
\thanks{Zhijin Qin and Xiaoming Tao are with the Department of Electronic Engineering, Tsinghua University, Beijing 100084, China. (email: qinzhijin@tsinghua.edu.cn, taoxm@tsinghua.edu.cn)}
\thanks{Geoffrey Ye Li is with Imperial College London, London SW7 2AZ, U.K. (e-mail: geoffrey.li@imperial.ac.uk)}}

\maketitle
\begin{abstract}
The metaverse is expected to provide immersive entertainment, education, and business applications. However, virtual reality (VR) transmission over wireless networks is data- and computation-intensive, making it critical to introduce novel solutions that meet stringent quality-of-service requirements. With recent advances in edge intelligence and deep learning, we have developed a novel multi-view synthesizing framework that can efficiently provide computation, storage, and communication resources for wireless content delivery in the metaverse.
We propose a three-dimensional (3D)-aware generative model that uses collections of single-view images. These single-view images are transmitted to a group of users with overlapping fields of view, which avoids massive content transmission compared to transmitting tiles or whole 3D models. We then present a federated learning approach to guarantee an efficient learning process. The training performance can be improved by characterizing the vertical and horizontal data samples with a large latent feature space, while low-latency communication can be achieved with a reduced number of transmitted parameters during federated learning.
We also propose a federated transfer learning framework to enable fast domain adaptation to different target domains. Simulation results have demonstrated the effectiveness of our proposed federated multi-view synthesizing framework for VR content delivery.
\end{abstract}
\begin{IEEEkeywords}
Metaverse, virtual reality, multi-view synthesizing, federated learning, deep learning.
\end{IEEEkeywords}

\section{Introduction}
The emerging virtual interactions are offered through immersive three-dimensional (3D) experiences with holographic-type communications. Virtual reality (VR) applications have attracted massive attention for their revolutionary virtual and immersive user experience, offering new opportunities in various fields. The widely deployed devices require the next-generation wireless network to provide ubiquitous high-quality wireless services. Unlike conventional wireless services, a wireless VR system requires delivering tremendous high-resolution visual field resources with ultra-low latency. Additionally, corresponding content should be presented for all users with different viewpoints to avoid dizziness and nausea \cite{9430902}. Multiple gigabits per second per user and one terabyte per second aggregated are expected to enable immersive user experiences, such as VR video transmission and VR streaming \cite{9711518}.

Existing wireless technologies often struggle to meet quality-of-service (QoS) requirements for VR applications. In traditional video transmission schemes, videos are segmented into small chunks, which facilitates flexible resource allocation in wireless transmission and promotes high bandwidth efficiency \cite{li2022internet}. Given that a VR user has a limited field of view (FoV), roaming the entire 360-degree video frame is unnecessary. In tile-based VR content transmission, the chunks are further partitioned into rectangular tiles. This strategy enables multiple service qualities based on users' attention and viewports \cite{9133103}. Users requesting the same tile share a pilot in each multicast stream. However, the performance is limited by viewport prediction accuracy, and transmission latency is impacted by computational complexity. Additionally, the delivery of identical VR content varies due to a compound representation for users with different FoVs. Different tiles still need to be transmitted, especially when users are moving within the digital space. For example, in a VR streaming scenario, two users with avatars are located within a virtual room with one object situated in the center. Although both users observe the same object in the virtual room, the content delivered to each user is different as they perceive the object from different angles. The heterogeneity of requests for the same VR content is the primary distinction between VR content and traditional one, which poses a challenging problem in bandwidth resources for numerous users with diverse FoVs.

To provide richer, more engaging, and immersive experiences, significant research attention has been paid to improving the performance of wireless VR networks. Most existing VR content delivery methods explore the optimization algorithms in a specific wireless network \cite{9894271}. Even though high throughput can support VR applications, delivering VR content incurs a significant communication cost. We can borrow the idea of semantic communication \cite{10000901} to reduce the volume of required data and extend VR applications to larger-scale scenarios and various devices. Zhang \emph{et al}. \cite{zhang2023semantic} have proposed a semantic communication framework for extended reality with a universal variable-length semantic-channel coding method to adjust the coding process accordingly. In this method, semantically unimportant information is highly compressed or discarded.  Although data compression can mitigate the high bandwidth requirements for certain VR content, providing VR services for massive connections remains a challenging issue. As a result, exploiting the unique characteristics of VR users, who request different content from different viewpoints, has significant potential to save communication resources.

As an emerging framework, the neural radiance field (NeRF) has shown a powerful ability \cite{mildenhall2020nerf} for synthesizing novel views of complex scenes by optimizing an underlying continuous volumetric scene function. In NeRF, a bounded three-dimensional (3D) volume is used to define a learned, continuous volumetric radiance field to represent a scene. The self-supervised monocular scene reconstruction NeRF model developed by Cao \emph{et al}. \cite{cao2022scenerf} is trained from multiple image sequences with poses, geometry constraints, and a novel probabilistic sampling strategy. With the help of deep learning, it is possible to multicast the same VR content for users who request different views. To further improve communication efficiency, we consider VR scenarios implemented by stereoscopic imaging. This technique is used in VR to create a sense of depth and three-dimensionality by providing slightly different images to each eye, thus mimicking how humans perceive the world in reality. Despite the potential of generative models in VR communication, their practical application presents a significant challenge. To obtain a well-trained model, vast amounts of data are required. While generative models can conserve communication resources in VR scenarios, transmitting large volumes of unprocessed data to a central processor for training may lead to considerable traffic congestion and communication delays. 

Thus, centralized learning may be unsuitable for time-critical applications, where data is too large or costly to centralize. Furthermore, performing the entire training process centrally can impose significant computational overhead on the central processor, resulting in significant delays. Additionally, the training data from individual users might contain sensitive information, such as health and financial data \cite{cao2023communication}. Users who value their privacy may not be willing to share this data with others. These challenges motivate the emergence of federated learning. Unlike centralized methods, federated learning involves training machine learning models on decentralized devices by aggregating locally computed updates rather than raw data, thus preserving privacy and reducing data transmission overheads \cite{tao2023content}. However, frequently exchanging model parameters and waiting for aggregation to the next communication round results in poor communication efficiency and huge latency. Moreover, the heterogeneity from local datasets may introduce bias to the global model and slow down the convergence speed. Additionally, certain tasks in VR scenarios might have insufficient training data, resulting in performance degradation or even model collapse. For VR content, the data size needs to be large enough to obtain a well-trained model \cite{yang2019federated}. Federated transfer learning combines the principles of federated learning and transfer learning, allowing models to be trained on decentralized data sources while utilizing knowledge from pre-existing models, ensuring both data privacy and improved performance on local tasks.

Motivated by the aforementioned, we propose a 3D-aware generative model for wireless VR networks to deliver VR content efficiently. Compared to the tile-based content delivery scheme, the proposed model can multicast VR content to a set of users and generate multi-view consistent images based on their viewport. Note that the proposed model synthesizes novel views based on a single-view input, which may not yield images that precisely match the actual observed views of the object. However, this method reduces both bandwidth costs and latency for processing reusable content intended for multiple users. It can be particularly beneficial for VR scenarios requiring ultra-high reliability to prevent symptoms such as dizziness and nausea, as well as for resource-constrained VR applications. Considering the rising privacy concerns, users may be unwilling to share their viewing data for model training, and content providers are not allowed to share their users’ viewing history without permission. Additionally, exchanging raw data during training brings significant communication overhead. To address these challenges and leverage edge computational capabilities, we propose a federated learning scheme. This approach utilizes local dataset characteristics and employs pre-trained models across decentralized devices, benefiting from local data adaptations.
The main contributions of this paper are summarized as follows:
\begin{itemize}
  \item We propose a novel wireless VR content delivery network, in which the base station (BS) multicasts a sparse set of input views to a group of users with overlapped FoVs. The requested VR content is synthesized at the user side based on the user's viewport. We propose a multi-view synthesizing model, which can improve rendering efficiency and 3D consistency for the VR content delivery network.
  \item To efficiently train the multi-view synthesizing model, we develop a federated learning framework, which utilizes the characteristics of the local datasets, namely horizontal federated learning and vertical federated learning. Each user transmits fewer parameters but provides rich information to the global model, resulting in efficient training of the model.  
  \item To accelerate the training process for new tasks, we propose a federated transfer learning framework for the proposed VR content delivery network. A new loss function is designed to improve learning performance. The proposed federated transfer learning framework can efficiently accelerate domain adaptation and reuse the knowledge learned from the source model. 
\end{itemize}

\section{Related Works}
It can be observed that most of the current works on VR content delivery focus on improving the throughput of wireless VR networks. Chen \emph{et al}. \cite{9789210} have designed an iterative algorithm that iteratively optimizes the truncated first-order Taylor approximation of the objective for wireless multiplayer interactive VR game transmission framework based on mobile edge computing. Liu \emph{et al}. \cite{9565222} have developed a constrained deep reinforcement learning algorithm to select the optimal phase shifts of the reconfigurable intelligent surface under latency constraints. Gupta \emph{et al}. \cite{gupta2022mmwave} have proposed geometric programming algorithms and an intermediate step of graph-theoretic VR user to mmWave access point assignment. Huynh \emph{et al}. \cite{van2022edge} have addressed the optimization of the latency/reliability in digital twins-enabled metaverse by optimizing offloading portions, edge caching policies, bandwidth allocation, transmit power, computation resources of user devices, and edge servers. Song \emph{et al}. \cite{dang2022low} have proposed a way to allocate resources in multiple dimensions based on the quality of experience (QoE) and reduce the multi-cell interference for wireless mixed reality in dynamic time division duplex networks. Chaccour \emph{et al}. \cite{chaccour2022can} have conducted a quantification of the risk for an unreliable VR performance through a novel and rigorous characterization of the tail of the end-to-end delay. Li \emph{et al}. \cite{10124955} have proposed a deep reinforcement learning network to cache popular VR video chunks proactively. In general, existing works on VR applications mainly focus on reducing transmission delay via caching and wireless resource allocation for pre-stored video resources. However, the main characteristic of VR content delivery still lies in transmitting different content to users requesting the same video, which consumes vast communication resources.

Some works utilize the reusable characteristics of the shared FoVs among users. Users sharing highly overlapped FoVs can be divided into smaller groups and receive narrower frames or tiles. The deep reinforcement learning algorithm proposed by Zhang \emph{et al}. \cite{9567690} studies the viewport prediction data rate requesting problems. The scheme proposed by Liu \emph{et al}. \cite{liu2023rendered} reuses previously rendered tiles based on the prediction of the user's FoV, which leverages mobile edge computing to store and transmit the rendered tiles to the VR headset. The system presented by Jakob \emph{et al}. \cite{struye2022covrage} uses millimeter-wave beamforming and exploits the spatial and temporal correlation of the user's head movement to predict the optimal beam direction and switch between beams quickly. Tile-based VR streaming still consumes more wireless resources than traditional content delivery schemes since users always request different tiles to experience omnidirectional visual content. Additionally, the performance of tile-based VR content transmission highly relies on viewport prediction accuracy \cite{9798771}. The re-transmission of VR content can cause significant communication latency. 

The development of deep learning networks has enabled VR video streaming by multicasting single-view images instead of entire three degrees of freedom VR content. The ability to recover 3D shapes from single-view images in an unsupervised manner avoids massive data volume and bandwidth consumption. The NeRF model allows VR content delivery to be achieved in a pixel-wise manner, avoiding the need to roam the entire 3D model. Wang \emph{et al}. \cite{wang20224k} have explored ray correlation to enhance high-frequency details using geometry-aware local context to pursue high-fidelity view synthesis on ultra-high resolutions. Li \emph{et al}. \cite{li2022compressing} have designed a robust and adaptive metric for estimating redundancy and trainable vector quantization to improve the compactness of grid models. In combination with a joint tuning strategy and post-processing, the proposed method can achieve a high compression ratio. Additionally, generative adversarial networks (GAN) have recently extended their capabilities to 3D settings, which has gained momentum as well \cite{gu2021stylenerf}. Pan \emph{et al}. \cite{pan2020gan2shape} have mined 3D geometric cues from an off-the-shelf two-dimensional (2D) GAN that is trained on RGB images only, demonstrating that such a pre-trained GAN contains rich 3D knowledge. Jeong \emph{et al}. \cite{kwak2022injecting} have proposed a 3D-aware GAN that can discover semantic attributes and controls in an unsupervised manner. Cai \emph{et al}. \cite{cai2022pix2nerf} has used a pipeline to generate NeRF of an object or a scene of a specific class from a single-view image without 3D, multi-view, or pose supervision.

To the best of our knowledge, the unique approach of using a federated multi-view synthesizing model for VR content delivery, which takes into account the unique characteristics of VR requests, has not yet been explored in the context of wireless communication. There is a compelling need in the field to integrate the novel view synthesizing capabilities of NeRF into the challenges posed by wireless VR services.

The rest of this paper is organized as follows. The system model of the proposed VR transmission network and federated training are introduced in Section III. Section IV presents the proposed federated multi-view synthesizing model with vertical and horizontal federated learning. Moreover, we present a federated transfer learning scheme to accelerate the training process for new tasks. We provide the simulation results in Section V and conclude the paper in Section VI.

\begin{table}[!t]
    \centering
    \caption{List of Key Notations}
    \begin{tabular}{|c|c|}
        \hline
        Notation & Definition \\
        \hline
        $\mathcal{K}$ & Set of users \\ 
        $\mathcal{G}$ & Set of groups \\ 
        $\mathcal{C}$ & Set of clients \\ 
        $L_r$ & Pre-rendering latency \\
        $L_c$ & Content transmission latency \\
        $L_p$ & Post-processing latency\\
        $L$ & Total VR content delivery latency \\
        $S_k$ & Local dataset of user $k$ \\ 
        $S^h$ & Local horizontal federated datasets \\ 
        $S^v$ & Local vertical federated datasets \\
        $S_{tr}$ & Target domain datasets \\
        $H$ & Feature space \\
        $V$ & Sample ID space \\
        $p$ &  Coordinate positions of an object \\ 
        $d$ &  View directions from a camera \\ 
        $\zeta$ &  Ray-traced pixel\\
        $w$ &  Latent vectors \\
        $T$ &  Global communication round numbers\\    
        $T_r$ &  Index of global communication\\
        $\phi$ & Parameters in discriminators \\
        $\theta$ & Parameters in generators \\ 
        $G_{\theta}$ & Generators with $\theta$ \\ 
        $D_{\phi}$ & Discriminators with $\phi$ \\
        $\mathcal{L}(G_{\theta},D_\phi)$ &  GAN loss function \\ 
        $\mathcal{L}_v$ &  Federated vertical generator loss \\ 
        $\mathcal{L}_s$ &  Divergence of internal distribution loss\\
        $\mathcal{L}_g$ &  Geometry loss \\ 
        $\mathcal{L}_v$ &  Image quality loss\\
        $\mathcal{L}_{tr}$ &  Federated transfer learning loss function \\ 
        $\hat{G^s_{\theta}}$ & Generators with frozen layers \\
        \hline
    \end{tabular}
    \label{notations}
\end{table}

\section{System Model}
In this section, we introduce the system model, including general VR content delivery workflow, proposed VR network, and federated learning scheme.

\subsection{Problem Statement}
Consider a deep learning-enabled VR network, which includes a BS as the VR content provider and viewers as users, $\mathcal{K} = \{ 1,...,K\}$. The users can be clustered into $\mathcal{G} = \{ 1,...,G\}$ groups based on overlapped FoVs. The BS is equipped with an edge computing server, which enables the BS with the ability of computation and storage. The BS provides viewers VR streaming and video services over the wireless link. We present a workflow for the multi-view synthesizing VR content delivery system to further explain the issue. At the time slot $t$, the MEC server assigns rendering resources $r_k$ for viewer $k$ and generates the FOV raw image data using the latest VR content. Once the rendering process is complete, the FOV raw image data is transmitted to users for post-processing so that it can be adapted to VR display. The VR content required by user $k$ is represented as multiple viewports of an object. Then, signal $y_k$ received by user $k$ can be given as
\begin{equation}\label{y}
y_k=x_kh_k+ \sigma_k,
\end{equation}
where $x_k$ denotes the requested VR content for user $k$, $h_k$ denotes the channel between the user $k$ and the
BS, and $\sigma_k$ denotes the additive white Gaussian noise (AWGN), i.e., $\sigma_k \sim \mathcal{C}\mathcal{N}(0,\sigma^2)$. We assume a block-fading channel model that remains invariant in each coherence interval. 

The proposed VR system aims to provide high-quality VR content based on a selection of FoVs with high multi-view consistency, while also reducing latency and bandwidth costs. The latency comprises three parts: pre-rendering latency, transmission latency, and post-processing latency. As the number of VR devices can be numerous, the available CPU capacity of the edge server $R_c$ may not be sufficient. When the VR content is represented in a sphere space, the edge server needs to render the whole 3D Fov or 2D FoVs with heterogeneity for every user. The pre-rendering latency for all users at the $t$-th time slot to render a standard-size FOV raw frame image can be calculated by
\begin{equation}\label{latencypre}
L_r(t)=\frac{\sum_{k}^{K}\lambda x_k(t)}{R_c(t)},
\end{equation}
where $\lambda$ denotes the cycles needed to post-process every bit of data. For the transmission latency, we consider the frequency division duplex (FDD) in the system. Then, the transmission latency at $t$-th time slot for all users can be obtained by
\begin{equation}\label{latencytrans}
L_t(t)=\sum_{k}^{K}\frac{n^x_k(t)}{B_k(t)R_k(t)},
\end{equation}
where $n^x_k$ denotes the size of the transmitted signal for user $k$, and $B$ denotes the bandwidth allocated to VR users, and $R_k = log2(1+P_kh_k/\sigma^2)$ denotes the data transmission rate for given transmit power $P_k$, and noise power $\sigma^2$ denotes the noise power. The post-rendering latency, $L_p$, depends on the local computing capacity and post-processing can be done simultaneously on each VR device. In this paper, we assume that each VR device has the same computation ability for simplicity. Then, the total latency for VR content transmission in the system is 
\begin{equation}\label{latency}
L=L_r(t)+L_t(t)+L_p(t).
\end{equation}

Due to the limited user movement range, users can acquire immersive and interactive experiences by obtaining a few viewports of the VR content. The viewport movement of VR users at the $t$-th time slot can be modeled by an independent Gaussian distribution with variance $e(t)$ and zero mean $\mathcal{N}(0,e(t))$. Obtaining a well-trained multi-view synthesizing model requires a large amount of data, while each user only has limited data for training. As a result, it is hard to obtain a well-trained stand-alone model \cite{acar2020federated}. To obtain a generalized model with an underlying distribution of the overall dataset and decrease the energy, spectrum, and computational resource consumption, the federated learning approach is applied in our system. The model parameters $\theta$ are obtained by minimizing a certain loss function based on the whole data set from all users. 

The general wireless federated learning uses an iterative approach to train the model. 
In each iteration, a subset $\mathcal{C} \in \mathcal{K}$ is selected as the client to perform the training process.
The clients train multi-view synthesizing models by their local datasets. Each client trains the multi-view synthesizing model for one or a few epochs per iteration and then updates the model parameters to the central server. The updated models are aggregated at the center. The BS multicasts the aggregated model to all clients for the next stage of training. Compared with centralized training, which exchanges raw data, federated learning significantly reduces the communication overhead as fewer bits are transmitted to enable online training. Note that there is another popular distributed machine learning approach, named split learning. Split learning divides the full machine learning model into multiple smaller network portions and trains them separately, some parts on a server and others on distributed clients \cite{thapa2022splitfed}. On the other hand, federated learning runs a complete network on the client side. As our proposed VR content delivery scheme requires a well-trained model to synthesize novel views with multi-view consistency, a computationally heavy task, we apply the federated learning framework rather than the split learning framework.

After any iteration, the center shares the updated model with users. Then, the objective of federated learning is to achieve a global goal under the coordination of the center. The proposed VR content delivery scheme is utilized by designing a federated multi-view synthesizing model, and our focus lies in reducing the communication cost in the proposed VR network.

\subsection{Proposed VR Network}

\begin{figure}[t]
\centering
\includegraphics[width =\linewidth]{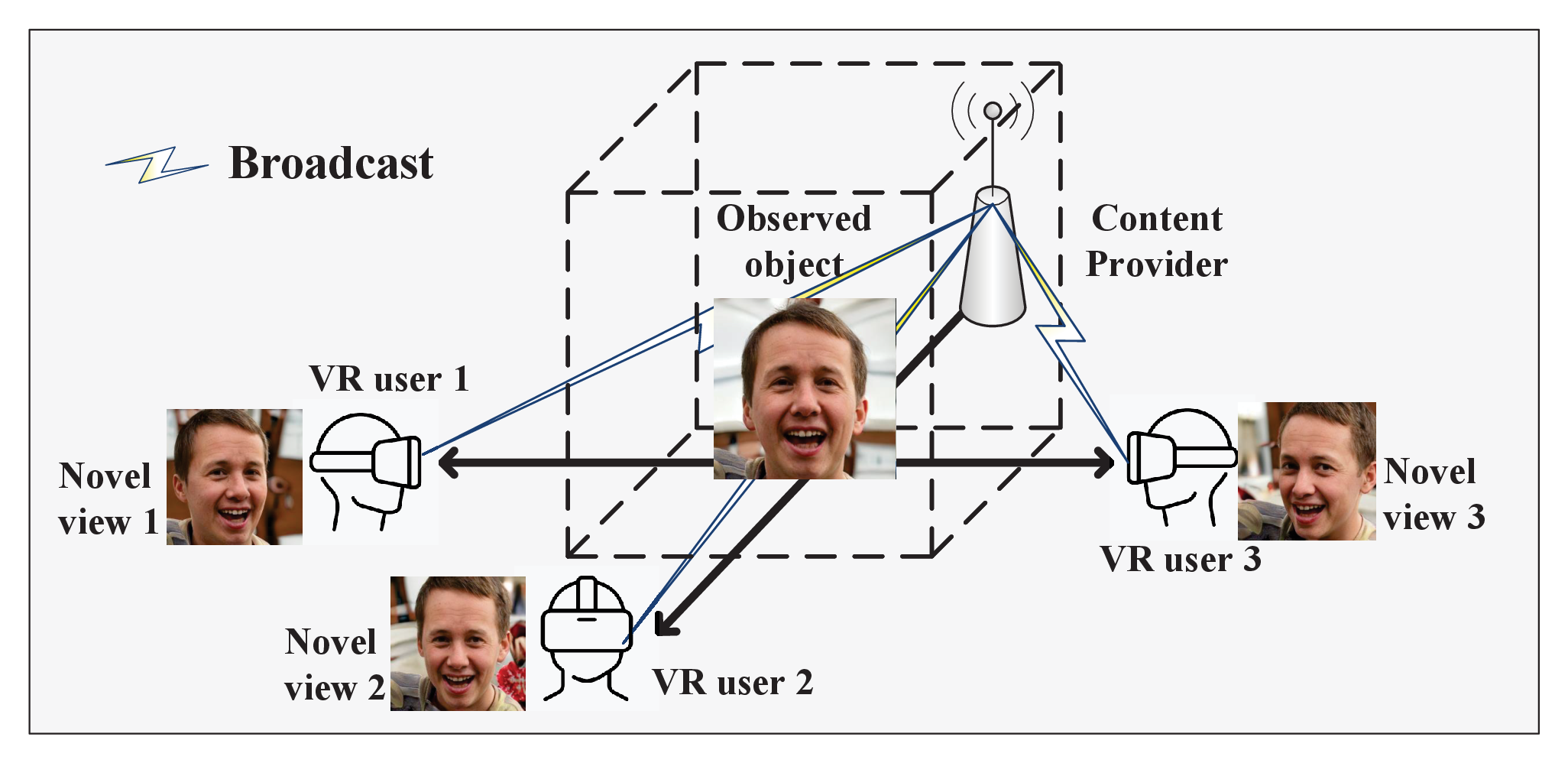}
\caption{Proposed wireless VR scheme, where the content provider multicast the single-view input to users; Users apply the 3D-aware generative model to synthesize required VR content by viewports.}
\label{illl}
\end{figure}

In the proposed transmission scheme, the VR content provider only multicasts a single-view image to a user group $g \in \mathcal{G}$. Then, the users in $g$ can synthesize the required viewports by a well-trained generative model, as shown in Fig. \ref{illl}. It can be seen that VR content at $t$-th time slot $x(t)$ has less common frame information or shared tiled for VR users. Thus, the traditional tile-based VR content delivery scheme still needs to transmit many different frames to different users. Denote $x_g$ as the single-view VR content multicast to group $g$ and $k_g$ users in group $g$. Then the pre-rendering latency in \eqref{latencypre} can be reduced by rendering one image for multiple users as
\begin{equation}\label{newlatencypre}
L_{r}^{*}(t)=\frac{\sum_{g}^{G}\lambda x_g(t)}{R_c}.
\end{equation}
Likewise, the transmission latency in \eqref{latencytrans} can be formulated as 
\begin{equation}\label{newlatencytrans}
L_{t}^{*}(t)=\sum_{g}^{G}\frac{n^x_g(t)}{B_g(t)R_g(t)},
\end{equation}
where $R_g$ denotes the transmitting rate of the VR user with the worst channel condition in group $g$. Then, $k_g$ use received $x_g$ to synthesize multi-view consistent VR content based on their viewports by a 3D-aware generative model, with the total latency represented by
\begin{equation}\label{latency}
L=L_{r}^{*}(t)+L_{t}^{*}(t)+L_p(t).
\end{equation}

In contrast to 2D generative models, 3D-aware generative models rely on the combination of a 3D-structure-aware inductive bias in the generator network architecture and a neural rendering network that aims at providing view-consistent results. The generative models are trained locally by the user viewing history as datasets. Each user has a local dataset $S_k = \{H_k, V_k \}$. where $H_k$ denotes feature space and $V_k$ denotes data identity document (ID) space. Each data sample is represented by a 2D image in the RGB model. 

With the limited data size that the local datasets have, the proposed generative model is designed to gain the ability of unsupervised shape generation. Compared with the existing tile-based VR networks, the proposed VR network can provide services to users with a range of FoVs. 
Unlike tile-based VR content delivery schemes, which constantly need to transmit different tiles from the captured video frames to all users, the proposed network shares transmitted images among a subset of users, thereby reducing bandwidth costs and transmission latency.

\section{Multi-view Synthesizing based on Federated Learning}
In this section, we propose a NeRF-based model to perform multi-view synthesizing, which enables the proposed VR transmission scheme. Then, we apply federated learning to train the model efficiently. We divide the datasets as vertical and horizontal datasets to further reduce latency in the process of federated learning.
\subsection{Multi-view Synthesizing Model}
Based on recent studies, 3D scenes can be represented as neural implicit fields. 
NeRF and multiplane images (MPIs) utilize volumetric neural implicit fields to synthesize novel scene representations.
The volumetric representation in MPIs is discrete and permits fast rendering, whereas NeRF and its variants use a continuous spatial representation\cite{zhao2022generative}. The NeRF is typically parameterized as multilayer perceptrons (MLPs) given by 
\begin{equation}\label{Nerf}
\begin{aligned}[b]
\Gamma^L(p, d) = &[\sin{2^0(p,d)},\cos{2^0(p,d)}, ..., \\ 
& \sin{2^{L-1}(p,d)},\cos{2^{L-1}(p,d)}],
\end{aligned}
\end{equation}
where $\Gamma$ is a mapping from $\mathbb{R}$ into a higher dimensional space $\mathbb{R}^{2L}$, $p = (p_1, p_2, p_3)\in \mathbb{R}^3$ denotes coordinate position, and $d = (d_1, d_2) \in \mathbb{R}^2$ denotes view direction. To synthesize an image $I \in \mathbb{R}^{H \times W \times 3}$, it renders each pixel independently by casting a ray that can be written as
\begin{equation}\label{ray}
r(q) = o + qd,
\end{equation}
where $o \in \mathbb{R}^3$ denotes the camera origin, and $q$ denotes the accumulated transmittance along the ray. 
Then, it aggregates color values along with their corresponding
densities.
Such a method achieves good performance but with a high computational cost. 
All sampled points on the ray in NeRF need to call the network multiple times. 
The above methods require posed multi-view images or 3D mesh. 
As the VR content on the user side is 2D FoV, the whole 3D mesh is unnecessary, especially for scenarios where VR content changes rapidly. 
Thus, transmitting single-view images instead of multi-view images can benefit bandwidth efficiency. 
However, it also brings increasing complexity and 3D inconsistency. 

To overcome the challenge, we propose a synthesizing model that realizes volumetric rendering in 2D planes instead of geometry representations.
The proposed synthesizing model contains two networks, namely, a mapping network that transforms an initial, normally distributed latent to an intermediate latent code $w \sim \omega$ and a synthesis network $G$ that takes the latent vector $w$ to synthesize multiple feature planes to produce an output image from a noise vector $z$. For 3D-aware image generation, the generator extracts density and color features into $\omega$. Ray-traced pixel $\zeta$ can be modeled as
\begin{equation}\label{raytrace}
\zeta(r) = \int_{0}^{+\infty} Q(q) c_q(r(q),d)dq,
\end{equation}
where $c_q(r(q),d)$ denotes the predicted RGB data from a weighted sum of all sampled colors by the latent vector $w$, i.e., $c_q(r(q),d) = l_c \circ w$, where $l_c$ denotes a 2-layer MLP. The accumulated transmittance is represented by $Q(q)$ as
\begin{equation}\label{transmittance}
Q(q) = exp(- \int_{0}^{q} \xi (r(s))dq)\xi (r(q)),
\end{equation}
where $\xi (r(q))$ denote the density obtains by the linear projection from \eqref{Nerf}. To further explore generative frameworks with partial 3D control over the underlying object in terms of texture/structure decomposition and novel view synthesis \cite{skorokhodov2022epigraf}, an intuitive way to train a 3D generative model is to train it with 3D data or corresponding latent space, which however requires explicit 3D supervision with a large volume of data and computational capability, and requires RGB datasets with segmentation masks, key points, or multiple viewports for an object. 
In the proposed VR content delivery network, each local dataset only contains VR frame RGB data. The proposed synthesizing model trains a NeRF-based generator from purely RGB data from scratch for the 3D structure extraction. 

The NeRF-based generation can be formulated in the GAN-based framework. 
GAN, first proposed in \cite{goodfellow2014generative}, has been widely applied in image processing and text generation. 
A typical GAN generally consists of a generative model $G_{\theta}$ and a discriminative model $D_{\phi}$, where $\theta$ and $\phi$ denote parameter sets of the generator and the discriminator, respectively. 
Both networks are trained simultaneously and try to compete with each other.  
With the help of the discriminator, the generative model can achieve the ability to synthesize novel samples similar to the target data.
The objective of a generator is to learn the distribution of target data while a discriminator is trained to distinguish the true samples from the synthesized samples. The loss function is shown as
\begin{equation}\label{GAN}
\begin{aligned}[b]
\min \limits_\theta \max \limits_\phi \mathcal{L}_{G} & (G_\theta,D_\phi)=
\mathbb{E}_{v_r}[log D_{\phi}(v_r)] \\
& +\mathbb{E}_{v_n}[log(1 - D_{\phi}(G_\theta(v_n))],
\end{aligned}
\end{equation}
where $\mathbb{E}$ is the expectation operator, $v_r$ denotes the real data, and $v_n$ denotes random noise input. Both generator and discriminator consecutively repeat the process of training, where $D_\phi$ is trained to increase the value when the input is from the real dataset and decrease the value when the input is from the generator. The generator is trained to fool the discriminator. Essentially, the generator and the discriminator are trained in an adversarial way to achieve the Nash equilibrium. Given an optimal generator $G_\theta^{*}$, the inputs for the discriminator, $v_r$ and $G_\theta^{*}(v_n)$, have a similar distribution that the discriminator cannot distinguish generated data from real data.

The GAN structure enables the generative model to synthesize novel camera control based on single-view images compared with the NeRF-based structure. The proposed GAN-based model cannot synthesize a full 3D geometry as the NeRF model but only provides a range of FoV. As our objective is to reduce the transmission overhead in the VR network, the proposed GAN realizes high bandwidth efficiency. We employ the NeRF network as the backbone but do not utilize view direction conditioning as the NeRF. With only single-view data available, view direction conditioning worsens multi-view consistency in GAN. The proposed GAN adopts a non-saturating GAN objective with R1 regularization, which is given by 
\begin{equation}\label{cGAN}
\begin{aligned}[b]
\mathcal{L}_{GAN}(G_{\theta},D_\phi) = & \mathbb{E}_{z \sim Z, p \sim P}[f(D_{\phi}(G_{\theta}(z,p))] + \\
&\mathbb{E}_{ I \sim p_{d}}[f(-D_{\phi}(I))+ \lambda|| \nabla D_{\phi}(I)||^2],
\end{aligned}
\end{equation}
where $z$ denotes the unit Gaussian noise as input, $p$ denotes the data synthesized by the generator $G_{\theta}$, and $p_d$ denotes the distribution of real data. To improve the multi-view consistency, the regularization term, $\mathcal{L}_{Reg}$, is added by the corresponding rays of the pixels in the low-resolution image generated via NeRF and high-resolution output of the proposed model. The overall loss function is 
\begin{equation}\label{fGAN}
\mathcal{L}(G_{\theta},D_\phi) = \mathcal{L}_{GAN}(G_{\theta},D_\phi) + \beta \mathcal{L}_{Reg},
\end{equation}
where $\beta$ is the weight of the regularization term. The model is trained progressively from low to high resolution. The NeRF structure is utilized to generate a low-resolution feature map, and then we employ progressively upsampling into the required resolution. The feature map is interpolated using predefined bilinear interpolation filters and pixel shuffle upsamplers. Applying upsamplers might cause minor distortion in image quality, but it significantly reduces computational complexity due to the lower resolution of the feature maps. It is worth noting that upsamplers generally lead to multi-view inconsistency. This artifact can be mitigated by progressively integrating the upsamplers \cite{chan2022efficient}.

\subsection{Vertical and Horizontal Datasets in Federated Learning}
\begin{figure}[t]
\centering
\includegraphics[width =\linewidth]{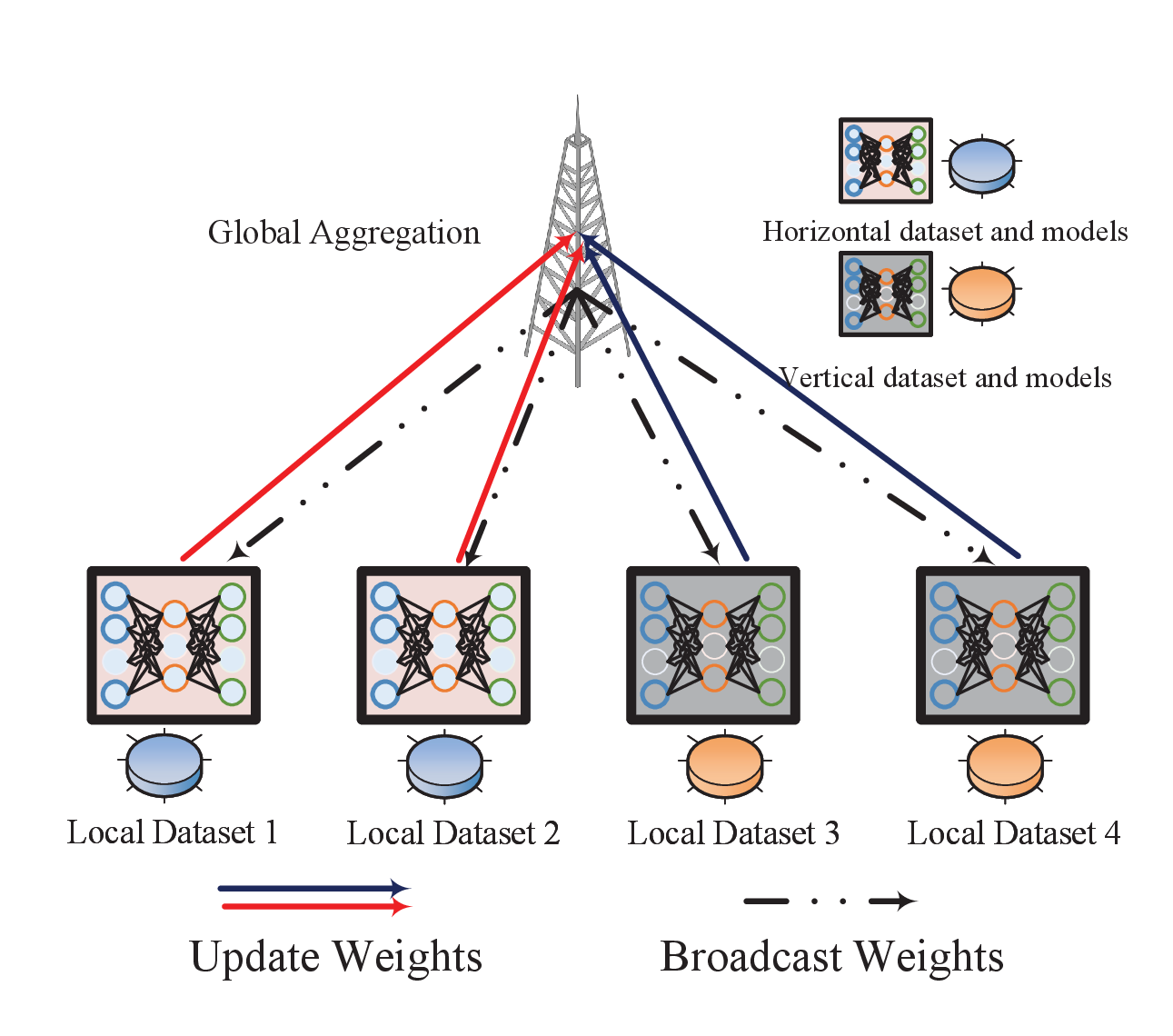}
\caption{Proposed federated learning for model training, where datasets are categorized to horizontal and vertical; Clients upload parts of the local models by the characterizing of their dataset.}
\label{fed}
\end{figure}
As aforementioned, the local datasets can be classified as horizontal and vertical datasets. 
Federated learning can be beneficial when a large number of clients participate. However, more clients can cause significant communication overhead when exchanging models. 
To reduce the communication overhead during the training and take advantage of the edge computing power, we propose a novel federated learning scheme based on the model structure shown in Fig. \ref{fed}. We assume the local users have the ability to carry out simple data analytics of their own datasets and to evaluate the quality of local models. Based on the evaluation, the local client will contribute the global model according to the data distributions of the local datasets.

In federated learning, each global iteration is called a communication round, $T_r \in T$. 
A communication round consists of local training, model updating, model aggregation, and model multicasting. 
Selected clients, $\mathcal{C}$, train the models locally with their local datasets $S_c \in \mathcal{S}$. 
As our target is to obtain a well-trained multi-view synthesizing model, the clients only upload the generators and center multicast aggregated generators to the clients. 
At the end of each communication round, all clients share the same global model. 
The local datasets are classified as horizontal and vertical datasets based on how data is distributed among various parties in the feature and sample space. 
Horizontal federated data samples are referred to as scenarios in which datasets share the same feature space but differ in sample space. In contrast, vertical federated data samples are for the cases where datasets share the same sample space but differ in feature space. 
We denote the horizontal federated dataset as $S_i^h = \{H_i^h, V_i^h\}, i\in C^h$, horizontal GAN parameter sets as $\theta_i$ and $\phi_i$, the vertical federated dataset as $S_j^v= \{H_j^v, V_j^v\}, j\in C^v$, and vertical GAN parameter sets as $\theta_j$ and $\phi_j$, where $C^h$ and $C^v$ denote the clients with horizontal and vertical datasets, respectively.

Note that the federated vertical datasets differ in ID space but still overlap with federated horizontal datasets in feature space. We neglect the overlapped features in federated vertical datasets and only apply the unique features that only federated vertical datasets have. To do so, we approximate \eqref{raytrace} as
\begin{equation}\label{approx}
i^{*}(r) = l_c \circ w \int_0^qQ(q)\Gamma^L(d)dq.
\end{equation}
Instead of calculating each ray and predicting the RGB data in \eqref{raytrace}, we aggregate 2D features space and then process a ray set as in \eqref{approx}. The object for the federated horizontal generator loss is given by $\mathcal{L}_h=G_h(\theta_i(w),\zeta(r))$.

We apply a hyper-network $G^*_v$ to increase the multi-view consistency for federated vertical datasets\cite{chan2021pi}, built upon \eqref{raytrace}. The object for the federated vertical generator loss is given by
\begin{equation}\label{vloss}
\mathcal{L}_v=\frac{1}{H_j^v*} \sum_{H_j^v*} G_v(\theta_j(w),\zeta(r))-G_v^*(\theta_j^*(w),\zeta^*(r)),
\end{equation}
where $H_j^v*$ denotes a sub-sampling of the featured pixels, $G_v$ denotes the federated vertical generator, and $\theta_j^*(w)$ denotes the weights of the hyper-network $G^*_v$. The hyper-network, $G^*_v$, provides multi-view supervision upsampled from a low-resolution mesh. Since the proposed VR network only requires standard-size FOV raw frame images, obtaining 2D multi-view consistency instead of 3D mesh features saves training time and post-process latency. 

\begin{algorithm}[!t]
\caption{Proposed federated multi-view synthesizing model training.}
\begin{algorithmic}[1]
\STATE \textbf{Initialize:} VR content viewers $\mathcal{K}$, communication round $T_r$, dataset $\mathcal{S}$, local models $\theta_k$, and $\phi_k$, and global model $\theta$;
\STATE \textbf{Preprocess:} Local datasets are classified as vertical and horizontal datasets, horizontal datasets $S_h \in \mathcal{S}$, vertical datasets $S_v \in \mathcal{S}$;
\FOR{$T_r = 1,...,T$}
\STATE Selection of clients $\mathcal{C}$ form $\mathcal{K}$ ; 
\STATE Center multicasts $\theta(T_r-1)$ to clients;
\STATE In communication round $T_r$, parallel for $\mathcal{C}$:
\FOR{\text{epochs} $e$}
\STATE Each local model is trained by \eqref{fGAN} with local dataset $S_c$ and $\theta(T_r-1)$;
\STATE Clients upload the parts of their parameters $\theta_i(T_r), i \in C^h$ and $\theta_j(T_r), j \in C^v$;
\STATE The center aggregate $\theta^t$ by \eqref{fedavgw} and obtain $\theta_{ema}(T_r)$ by \eqref{ema};
\STATE $\theta(T_r-1) = \theta(T_r)$;
\ENDFOR
\STATE $T_r = T_r + 1$;
\ENDFOR
\end{algorithmic}
\end{algorithm}

The local models can only update parts of their parameters depending on the type of their datasets. For horizontal federated datasets with the same feature space, the specific layers are uploaded, including geometry synthesis, rendering, and color prediction. For vertical federated datasets with different feature spaces, the shared layers for object generation are uploaded, including viewpoint analysis. By doing so, each client can transmit models with a reduced amount of parameters. Although the proposed federated learning requires more clients to join the training, the overall latency can be largely reduced. Besides, the accelerated convergence speed leads to fewer communication rounds, which can further reduce the communication overhead. Our goal is to train the model using federated learning with improved convergence speed and model robustness. The global loss function is
\begin{equation}\label{fedavg}
\mathcal{L}(\theta) = \frac{\sum_{c=1}^C |S_c|\mathcal{L}(\theta_c)}{\sum_{c=1}^C |S_c|} ,
\end{equation}
where $\theta_c$ denotes the corresponding parameters of the generator for client $c \in \mathcal{C}$, and $\mathcal{L}(\theta)$ denotes the loss function with parameter set $\theta$. As the horizontal federated data samples share the same feature space, the performance of the global model is valued by the feature expressions from the horizontal federated data samples. The vertical federated data samples are trained to provide camera pose to ensure multi-view consistency. Note that the weights of the hyper-network, $\theta_j^*$, are kept on the client side to reduce the burden of communication. Denote $\theta_i(T_r-1)$ and $\theta_j(T_r-1)$ as the part of the generator for client $i \in C^h$, and the part of the generator for client $j \in C^v$, respectively. The center calculates $\theta(T_r)$ by weighted federated averaging given as:
\begin{equation}\label{fedavgw}
\theta(T_r) = \frac{\sum_{i=1}^{C^h} |S_i|\theta_i(T_r-1)}{\sum_{i=1}^{C^h} |S_i|} +
  \frac{\sum_{j=1}^{C^v} |S_j|\theta_j(T_r-1)}{\sum_{j=1}^{C^v} |S_j|}.
\end{equation}
The clients update their selected layers based on the dataset types. We apply the exponential moving average (EMA) at the aggregation stage. The server aggregates the uploaded model by
\begin{equation}\label{ema}
\theta_{ema}(T_r) = \gamma\theta_{ema}(T_r-1) + (1-\gamma)\theta(T_r),
\end{equation}
where $\theta_{ema}(T_r)$ denote the moving average variable in $T_r$ and $\gamma \in [0,1)$ denote the decay factor. As the discriminators are trained locally with bias, the aggregated generators may lead to poor performance. Updating the global model with EMA enables the aggregated parameters at the end of a communication round related to the previous communication rounds. The estimation of the mean value of the historical parameters offers robustness to the global model.

\subsection{Federated Transfer Learning}
The horizontal and vertical federated datasets may be ineffective for the marginal VR users, whose feature space and sample ID space have less overlap. Also, training a multi-view synthesizing model from scratch costs wireless communication resources greatly. To cope with this challenge, we can leverage the knowledge from the pre-trained synthesizing model and adapt it to a new domain to speed up training. Denote a group of marginal users as $g_{tr}$ and each contains a dataset $S_{tr} = \{H_{tr}, V_{tr} \}$. Group $g_{tr}$ has a different task and a target domain while can reuse the knowledge from a well-trained synthesizing model obtained from aforementioned horizontal and vertical federated learning datasets. The target domain satisfies the cross-domain correspondence, i.e, $\forall{\omega_{tr} \in \mathcal{W}_{tr}}$, $G_{\theta}(\omega)$ and $G_{\theta}(\omega_{tr})$ have similar similar VR content or shape, where $\mathcal{W}_{tr}$ denotes the latent space for $g_{tr}$. The GAN domain adaptation is usually achieved by transfer learning for both the generator and the discriminator. However, transfer learning using GAN loss may be unstable and cause serious model collapse. It can also be time-consuming as the pre-trained GAN model has achieved a Nash equilibrium. Applying different source domains requires the model to achieve a new balance first. 

Note that the generators from the pre-trained model can work alone to synthesize multi-view consistent results. The discriminators are more personalized and stored at the client side without aggregation. Thus, we apply the generator $G^s_{\theta}$ with the source domain to obtain an adapted generator $G^{tr}_{\theta}$ with the target domain. 
All clients in $g_{tr}$ receive the $G^s_{\theta}$ as the initial model, and their objective is to obtain $G^{tr}_{\theta}$ with domain adaptation. 
Considering the cross-domain between $\mathcal{W}_{tr}$ and $\mathcal{W}_{s}$, the geometry of the objects and camera pose layers are frozen. We denote the model with frozen layers as $\hat{G^s_{\theta}}$. Then each client in $g_{tr}$ will train $\mathcal{W}_{tr}$ from $\hat{G^s_{\theta}}$ with the local dataset. 

\begin{algorithm}[!t]
\caption{Proposed federated transfer multi-view synthesizing model training.}
\begin{algorithmic}[1]
\STATE \textbf{Initialize:} marginal VR content users $\mathcal{K}_{tr}$, communication round $t$, local datasets $\mathcal{S}^k_{tr}, k \in \mathcal{K}$, and source model obtained by federated learning $\theta_{s}$;
\STATE \textbf{Preprocess:} Camera and geometry layers are frozen, and the source domain model parameters are denoted as $\hat{\theta}_{s}$;
\FOR{$T_r = 1,...,T_0$}
\STATE Selection of clients $\mathcal{C}$ from $\mathcal{K}_{tr}$; 
\STATE Center multicasts $\hat{\theta}_{s}(T_r-1)$ to clients;
\STATE In communication round $T_r$, parallel for $\mathcal{C}$:
\FOR{\text{epochs} $e$}
\STATE Each local client trains $\hat{\theta}^c_{tr}$ by \eqref{mswd} with local dataset $S_{tr}^c, c \in \mathcal{C}$, and $\hat{\theta}_s(T_r-1)$;
\STATE Clients upload the parameters $\hat{\theta}^i_s(T_r), i \in \mathcal{C}^h$;
\STATE The center aggregate $\hat{\theta}_s(T_r)$ by \eqref{fedavg};
\STATE $\hat{\theta}_s(T_r-1) = \hat{\theta}_s(T_r)$;
\ENDFOR
\STATE $T_r = T_r + 1$;
\ENDFOR
\STATE Unfreeze all layers and fine-tune the rest parameters;
\FOR{$T_r = T_0,...,T$}
\STATE Selection of clients $\mathcal{C}$ from $\mathcal{K}_{tr}$; 
\STATE Center multicasts $\theta_{s}(T_r-1)$ to clients;
\STATE In communication round $T_r$, parallel for $\mathcal{C}$:
\FOR{\text{epochs} $e$}
\STATE Each local client trains $\theta^c_{tr}$ by \eqref{swd} with local dataset $S_{tr}^c, c \in \mathcal{C}$, and $\theta_s(T_r-1)$;
\STATE Clients upload the parameters $\theta^i_s(T_r), i \in \mathcal{C}^h$;
\STATE The center aggregate $\theta_s(T_r)$ by \eqref{fedavg};
\STATE $\theta_s(T_r-1) = \theta_s(T_r)$;
\ENDFOR
\STATE $T_r = T_r + 1$;
\ENDFOR
\end{algorithmic}
\end{algorithm}

To accelerate the training process and improve the model performance, we need to learn the target distribution from the source distribution offered by $\hat{G^s_{\theta}}$. We apply the slice Wasserstein distance (SWD) \cite{kolouri2018sliced} to minimize the divergence of the empirical internal distribution given as 
\begin{equation}\label{mswd}
\mathcal{L}_s = \frac{1}{n_s\| \Psi \|} \sum_{\psi \in \Psi} \sum_{n_s=1}Y(\psi(\hat{G^s_{\theta}}(\mathcal{W}_{tr})),\psi(\hat{G^s_{\theta}}(\mathcal{W}_{s}))),
\end{equation}
where $n_s$ denotes the number of the sample vectors from source latent space $\mathcal{W}_{s}$, and $\Psi$ denotes a set of convolutional layers from the source generator with the perceptual loss on texture. 
The SWD is given by  
\begin{equation}\label{swd}
Y(U, V) = \int_{s^C} || sort(pr(U, \delta)) -  sort(pr(V, \delta))||,
\end{equation}
where $s^C = \{ \delta \in \mathbb{R}^C: || \delta || = 1 \}$, $U, V \in \mathbb{R}^{H \times W \times C}$ denotes arbitrary tensor, $pr$ denotes a mapping from the $C$ channels to a scalar by $\delta$ for each pixel, and $sort$ denotes the ordering of values. This operation comes with a low computational complexity but can be achieved by convolution of a randomized $1 \times 1$ kernel. The internal distributions can provide rich features from the source domain \cite{zhang2022generalized}. For the geometry layers, we perform a similar operation to obtain the geometry loss as 
\begin{equation}\label{geo}
\begin{aligned}[b]
\mathcal{L}_g = \frac{1}{n_s\| \hat{\Psi} \|} \sum_{\psi \in \hat{\Psi}} \sum_{n_s=1}Y(& \psi(m_{n_s} \odot G^s_{\theta}(\mathcal{W}_{tr})),\\& \psi(M_{n_s} \odot G^s_{\theta}(\mathcal{W}_{s}))),
\end{aligned}
\end{equation}
where $\hat{\Psi}$ denotes a set of convolutional layers from the source generator with the perceptual loss on geometry, and $m$ and $M$ denote the upsampled samples synthesized from the source and target domains, respectively. To further narrow the gap between the source and target domains, image quality loss is added to guarantee visual consistency given by
\begin{equation}\label{tloos}
\begin{aligned}[b]
\mathcal{L}_i = & l_1(G^s_{\theta}(\mathcal{W}_{tr}),G^s_{\theta}(\mathcal{W}_{s})) + l_2(G^s_{\theta}(\mathcal{W}_{tr}),G^s_{\theta}(\mathcal{W}_{s})) \\& + l_3(G^s_{\theta}(\mathcal{W}_{tr}),G^s_{\theta}(\mathcal{W}_{s})),
\end{aligned}
\end{equation}
where $l_1$ denotes the negative structural similarity loss, $l_2$ denotes the perceptual loss, and $l_3$ denotes MSE loss. Then we have 
\begin{equation}\label{swdddd}
\mathcal{L}_{tr}=\lambda_1\mathcal{L}_s+\lambda_2\mathcal{L}_g+\lambda_3\mathcal{L}_i,
\end{equation}
where $\lambda_1$, $\lambda_2$, and $\lambda_3$ denote the weights of the corresponding loss. The overall algorithm for federated transfer learning is summarised in Algorithm 2.

\subsection{Complexity Analysis}
Our generator network is composed of multiple blocks, each consisting of several convolutional layers, adaptive instance normalization, and non-linear activations. Each block takes a lower-resolution feature map as input and produces a higher-resolution feature map as output, and the output from each block is combined with a learned skip connection. The generator has $N$ blocks, and the input noise vector has a dimensionality of $Z$. The number of floating-point operations required for a single forward pass of the generator is roughly proportional to $\mathcal{O}_G = N(C^2s_k^2 + 2HWC + C^2)log2(H),$ where C is the number of channels in each layer, $s_k$ is the size of the convolutional filter. $H$ and $W$ are the height and width of the output feature maps, respectively. Term $log_2(H)$ accounts for the upsampling operation in each block. 

The discriminator network is similar to the generator network, but it takes an image as input and produces a scalar value as output. The discriminator consists of multiple blocks, each consisting of several convolutional layers, instance normalization, and non-linear activations. Assume that the discriminator has $N^{*}$ blocks and the input image has dimensions of $C \times H \times W$. The number of floating-point operations required for a single forward pass of the discriminator is roughly proportional to $O_D = N^{*} (C^2s_K^2 + HWC) * log2(H)$. Overall, the time complexity of the proposed model is dominated by the generator network, which typically has more layers and higher-resolution feature maps than the discriminator. The exact time required for training or inference depends on the specific VR applications, the resolution of the generated images, the batch size used during training, and the specific hardware used. For instance, a VR task that demands high assurance against dizziness and nausea can be accommodated with a lower feature map dimension for $H \times W$. The desired output can be attained through upsampling, which not only ensures lower computational complexity and inference latency but also introduces some degree of distortion.

\section{Numerical Results}
In this section, we examine the proposed multi-view synthesizing model and federated learning scheme through numerical results. The simulation results evaluated that the proposed VR content delivery outperforms the conventional tile-based VR network in bandwidth cost and transmission latency when the VR users have fewer overlapped FoVs. The proposed federated learning scheme improves the synthesizing performance with lower communication rounds.

\subsection{Simulation Settings}
We consider a wireless edge-empowered VR network that consists of 20 VR users. The BS is located in the center and all VR users are located in a semi-circle area with an independent Gaussian distribution. The BS is with a gigabit bandwidth. Each user contains a partition of datasets. 
We utilize the Flickr-Faces-HQ (FFHQ) dataset for training and validation of the proposed VR content delivery scheme \cite{karras2019style}. The FFHQ dataset consists of 70,000 images of real human faces with a resolution of $1024^2$. We assume the human face to be captured at the origin point, with the pitch and yaw of the camera sampled from a Gaussian distribution. Furthermore, we employ the MetFaces dataset as the target domain for the federated transfer learning task \cite{karras2020training}. MetFaces is an image dataset comprising 1,336 PNG images of human faces extracted from works of art, each with a $1024^2$ resolution. Although the number of data samples limits the performance of multi-view synthesizing, both the FFHQ and MetFaces datasets contain images of human faces, providing shared knowledge that can be utilized to aid the training of the MetFaces dataset.

The FFHQ dataset is split into horizontal and vertical datasets by country. The horizontal datasets contain human face images from different countries, while the vertical datasets contain human face images from specific countries.

We inherit most of the hyperparameters from the StyleGAN2-ADA \cite{karras2020training}. In addition, the Fourier feature dimension is set to L = 10 for both fields. In \eqref{swdddd}, We set $\lambda_1= 5$, $\lambda_2= 0.2$, and $\lambda_3= 1$. We apply Fr\'echet Inception Distance (FID) \cite{heusel2017gans} and Kernel Inception Distance (KID) \cite{binkowski2018demystifying} as identified Metrics. FID denotes the distance measurement between the distributions of the features from images generated and features of real samples from the dataset. KID denotes the squared maximum mean discrepancy between inception representations. As there is no labeled data for novel views of human faces, the reference FID and KID are obtained from the overall validation datasets. Lower FID and KID denote better performance.

\subsection{Multi-view Synthesizing Model Performance and Visual Examples}

We first evaluate the performance of the proposed federated multi-view synthesizing model and compare the results with centralized training. Note that the input of the proposed synthesis network is not the in-the-wild images. but latent code generated by the mapping network with a camera pose. Taking in-the-wild images as input requires the GAN inversion to convert images into machine-readable content, which is not included in this work. We assume the VR content provider is able to provide content after inversion. 

\begin{figure}[t]
\centering
\includegraphics[width = \linewidth]{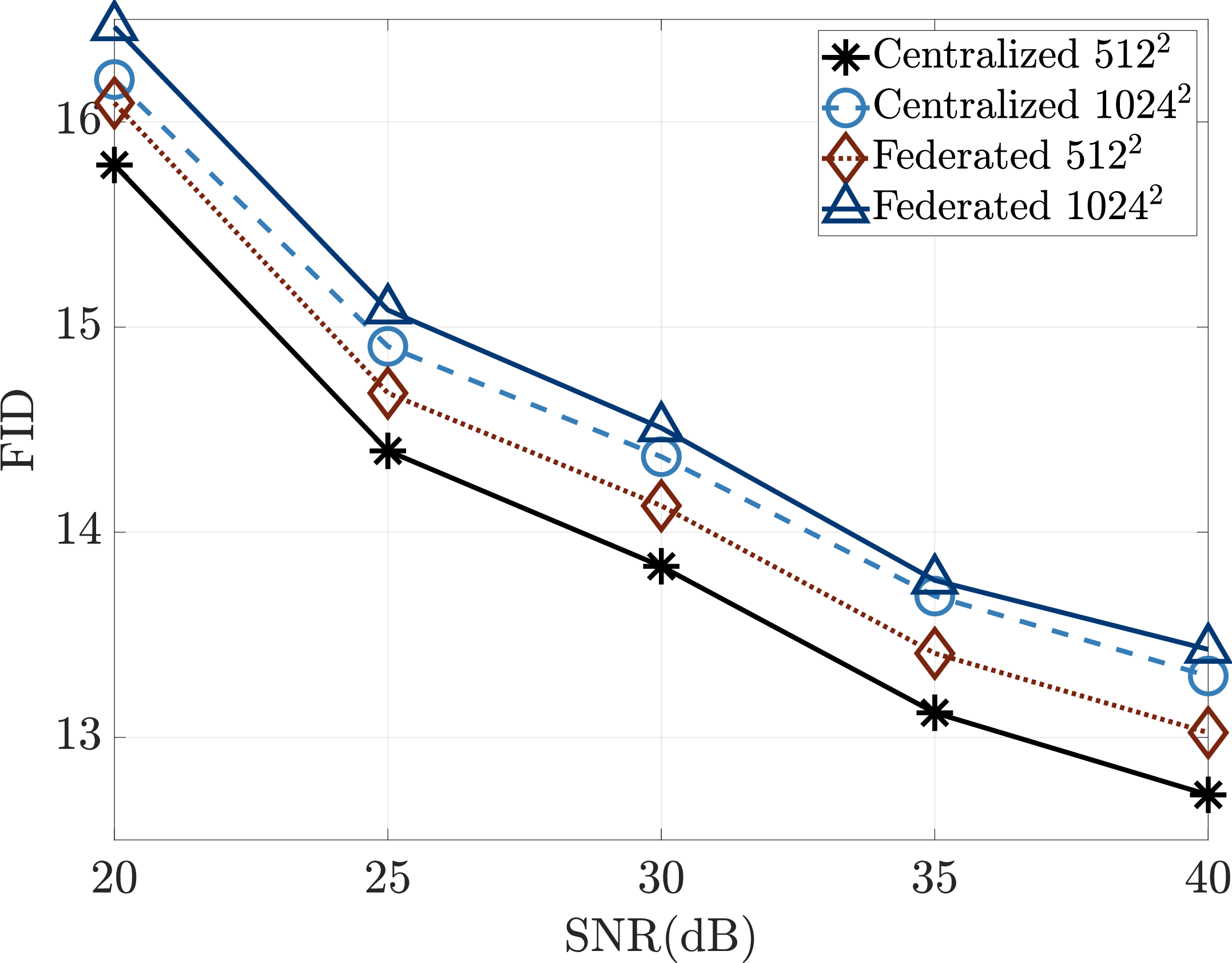}
\caption{FID performance of the proposed federated multi-view synthesizing model with different image SNR.}
\label{center}
\end{figure}
In Fig. \ref{center}, the proposed multi-view synthesizing model is evaluated under different image signal-to-noise ratios (SNR) in the VR content process for the VR transmission over wireless always experiences channel loss. Federated learning always has bias from the heterogeneity of the local models. Thus, we denote centralized training, which contains all data in the center, as the upper bound of the proposed multi-view synthesizing model. All VR users receive a frame with a resolution $1024^2$ or $512^2$ synthesized by the global model. The KID is applied as an indicator of the quality of service. From the figure, the performance of the proposed federated multi-view synthesizing model has an acceptable loss compared to the centralized training. In comparison, federated learning accesses the local datasets more timely and provides privacy protection. It also efficiently utilizes edge computing resources and avoids huge volumes of data exchange. It is easy to understand that higher resolution requires more features to reconstruct high-quality synthesized images. The performance will decrease as the resolution increases. High-quality content also leads to higher communication overhead during federated learning, as large feature maps cause increasing amounts of model parameters. Even so,
the proposed scheme maintains its advantages since the size of the local datasets will also increase dramatically. The proposed horizontal and vertical federated learning also decreases the required data overhead volume. 

\begin{figure}[t]
\centering
\includegraphics[width = \linewidth]{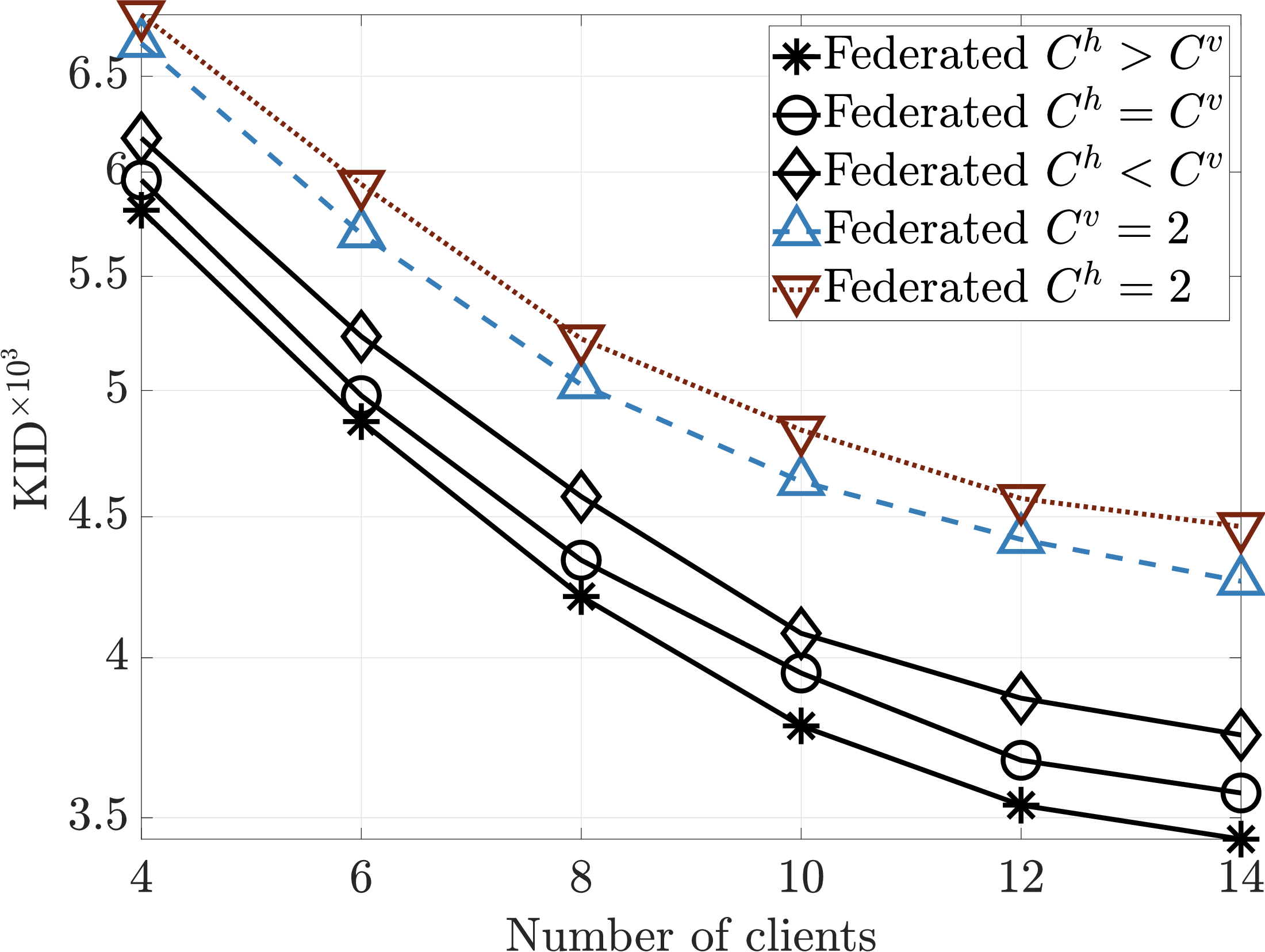}
\caption{KID performance of the proposed federated multi-view synthesizing model with different clients and dataset settings.}
\label{user}
\end{figure}
Fig. \ref{user} shows the performance of the proposed federated multi-view synthesizing model, which is tested with different numbers of clients. It can be seen that the participation of the clients benefits the performance. The horizontal datasets provide the layers that control volume rendering and feature expression while the vertical datasets provide the layers that control the camera pose. The horizontal datasets contain more diverse images and the vertical datasets have more bias as their feature spaces differ from the overall volumetric features. In particular, the fixed number of clients with horizontal datasets or vertical datasets results in poor performance. As the shortage of the number of clients who provide corresponding features, the global model will experience unbalanced aggregation. The horizontal datasets overlap less in ID space, i.e., countries, while overlapping a lot for human face feature expression. Thus, sufficient horizontal datasets can provide high-quality image generation. The vertical datasets are applied to offer camera pose and multi-view consistency as a feature supplement to analyze the NeRF of the human faces. It is reasonable to involve more horizontal datasets than vertical datasets to obtain solid volume rendering and camera pose. Moreover, the proposed federated learning scheme via datasets decreases the complexity compared to the federated learning scheme based on model segmentation.

\begin{figure}[t]
\centering
\includegraphics[width =0.93\linewidth]{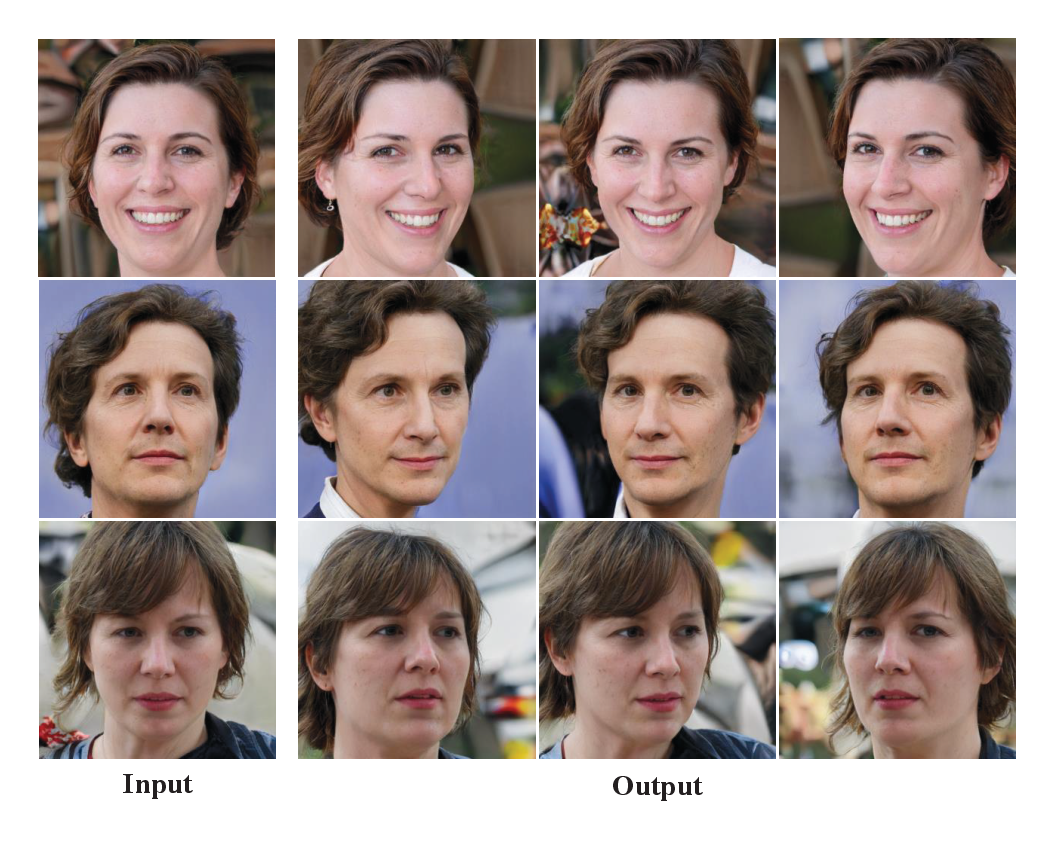}
\caption{Results for proposed VR network. With one single-view input, the users requesting a certain range of FoV can be served. It largely reduces the transmission overhead and latency, compared with the traditional VR schemes that have to transmit multi-view content separately. }
\label{Re1}
\end{figure}
As indicators, the FID and KID only reflect the quality of images, not the multi-view consistency. Some visual results are shown in Fig. \ref{Re1}. As we can see, the proposed model can provide consistent multi-view content from single-view input. The multi-view synthesizing avoids transmitting different content to a group of users who request the same content with different FoVs. The clients with vertical datasets are the providers of the multi-view layers. To ensure better performance of the multi-view consistency, the proposed federated learning framework adds the hyper-network to build the NeRF path from the vertical datasets. The proposed methods benefit the performance of the multi-view consistency but do not increase the communication overhead in federated learning. Only the generation network $G_v$ is updated and the hyper-network $G_v^{*}$ will not be transmitted during the model aggregation. The proposed VR content delivery scheme can multicast the full content to a group of users rather than tiles or only parts of the content. It decreases the bandwidth cost compared to traditional VR content delivery and provides a more immersive experience to VR users. The proposed model can also cover a range of FoVs, which requires no transmission when VR users are moving while the observed object is static. It should be mentioned that such a method trades extra computational resource consumption for communication and center pre-rendering resources. 
The proposed method guarantees multi-view consistency, rather than assures accuracy with higher fidelity. This is because the proposed model is trained in an unsupervised manner to improve communication efficiency during the training process. If a VR application prioritizes accuracy over communication efficiency, a supervised learning model could be applied in the proposed VR content delivery network to provide more accurate novel views by multicasting. Furthermore, the proposed model is more suitable for scenarios that demand ultra-reliability or communication resource-constrained scenarios with numerous VR users.

\subsection{Effectiveness in Latency}

\begin{figure}[t]
\centering
\includegraphics[width = \linewidth]{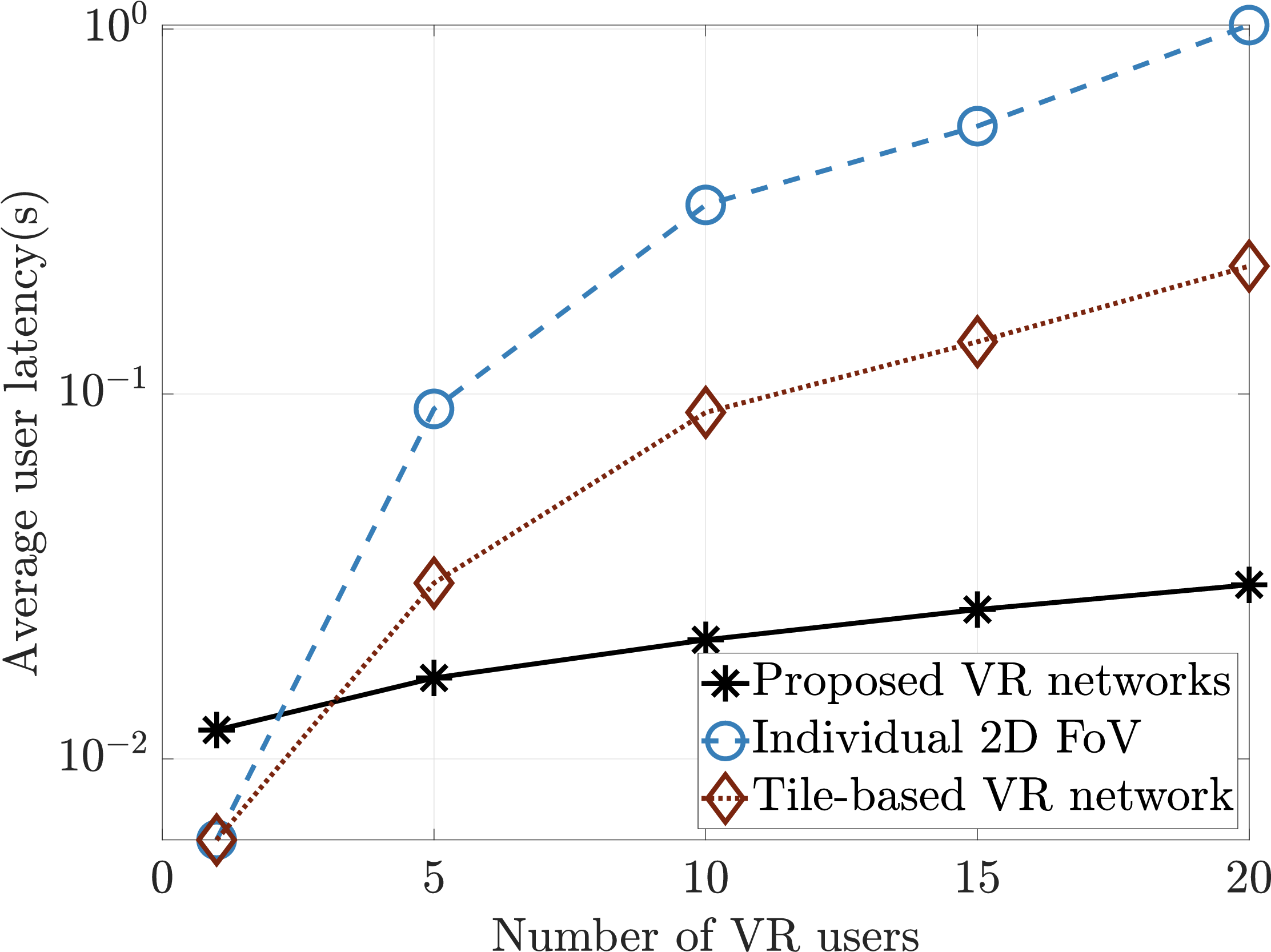}
\caption{Average latency of the proposed federated multi-view synthesizing network with different numbers of VR users.}
\label{late1}
\end{figure}
In Fig. \ref{late1}, we evaluate the proposed VR content delivery scheme from the latency aspect. 
We assume there are two groups if there are over two users, and each VR device has the same post-process computing capacity. 
We omit the latency caused by FoV prediction in tile-based VR networks and assume a fixed $20\%$ shared tiles \cite{perfecto2020taming}.
We consider a metaverse in that VR users have digital characters and space, which leads to less shared FoV or tiles. 
As aforementioned, the latency of the proposed system contains three parts: pre-rendering, communication, and post-process latency. 
The proposed method reduces the latency of both pre-rendering and communication. 
As only a single-view image needs to be pre-rendered, the content provider can utilize a higher computation efficiency for rendering. 
Multiple content rendering will cause computation resource shortage at the BS and lead to queuing when there are massive VR content requests.
The communication latency of the proposed VR content delivery method is only limited to the user with the worst channel condition. 
The proposed method experiences less performance loss when the number of VR users increases. The proposed multi-view synthesizing model can synthesize one image based on required viewports within 80 ms for a resolution of $512^2$ and 120 ms for a resolution of $1024^2$, using a single A5000 GPU. Although multi-view synthesizing brings extra latency for computing the novel 3D-aware content, synthesizing novel views can be done simultaneously, lowering the overall latency of the VR network.

\begin{figure}[t]
\centering
\includegraphics[width = \linewidth]{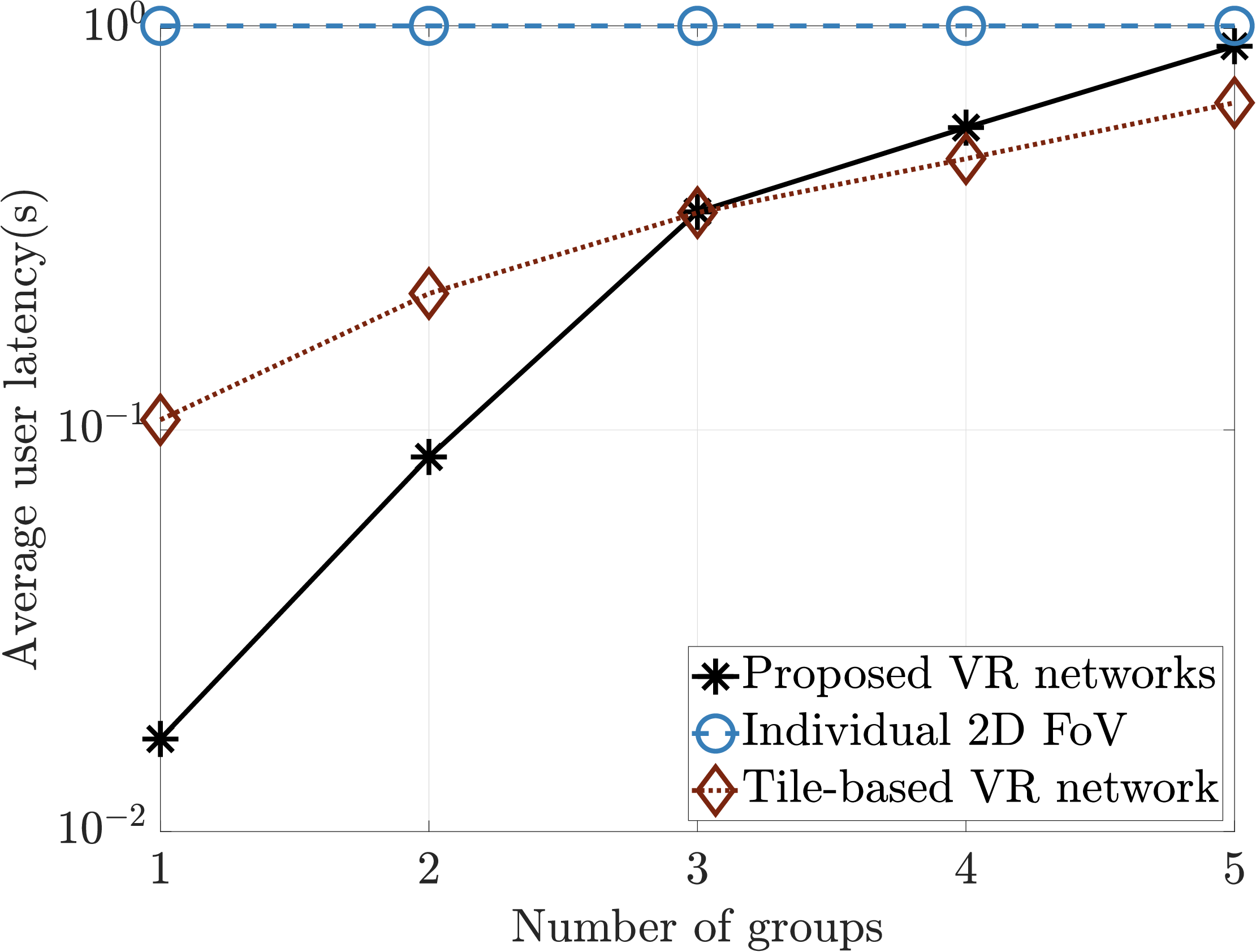}
\caption{Average latency of the proposed federated multi-view synthesizing network with different numbers of VR user groups.}
\label{late2}
\end{figure}
The latency performance with the number of groups is demonstrated in Fig. \ref{late2}, where 20 users are divided into groups by overlapped FoVs. The majority of the VR users are allocated to one group for the Gaussian distribution viewports. As the individual 2D FoV scheme needs to transmit unique VR content to different users, it cannot benefit from grouping and clustering. The tile-based network and the proposed multi-view synthesizing network both have performance gains when the number of groups is small. Serving fewer groups of VR users means more VR users have overlapped FoVs. The proposed VR content delivery scheme performs better in such scenarios and can support massive connections. However, when the number of groups increases, the proposed VR content delivery network suffers great performance loss. Since there are groups with only a few users, the proposed VR content delivery network can not utilize the single-view image efficiently but introduces extra computation latency to the users. This indicates that combining the proposed VR content delivery network with other networks, such as tile-based networks, can improve the overall latency performance.

\subsection{Ablation Study and Convergence}

\begin{figure}[t]
\centering
\includegraphics[width = \linewidth]{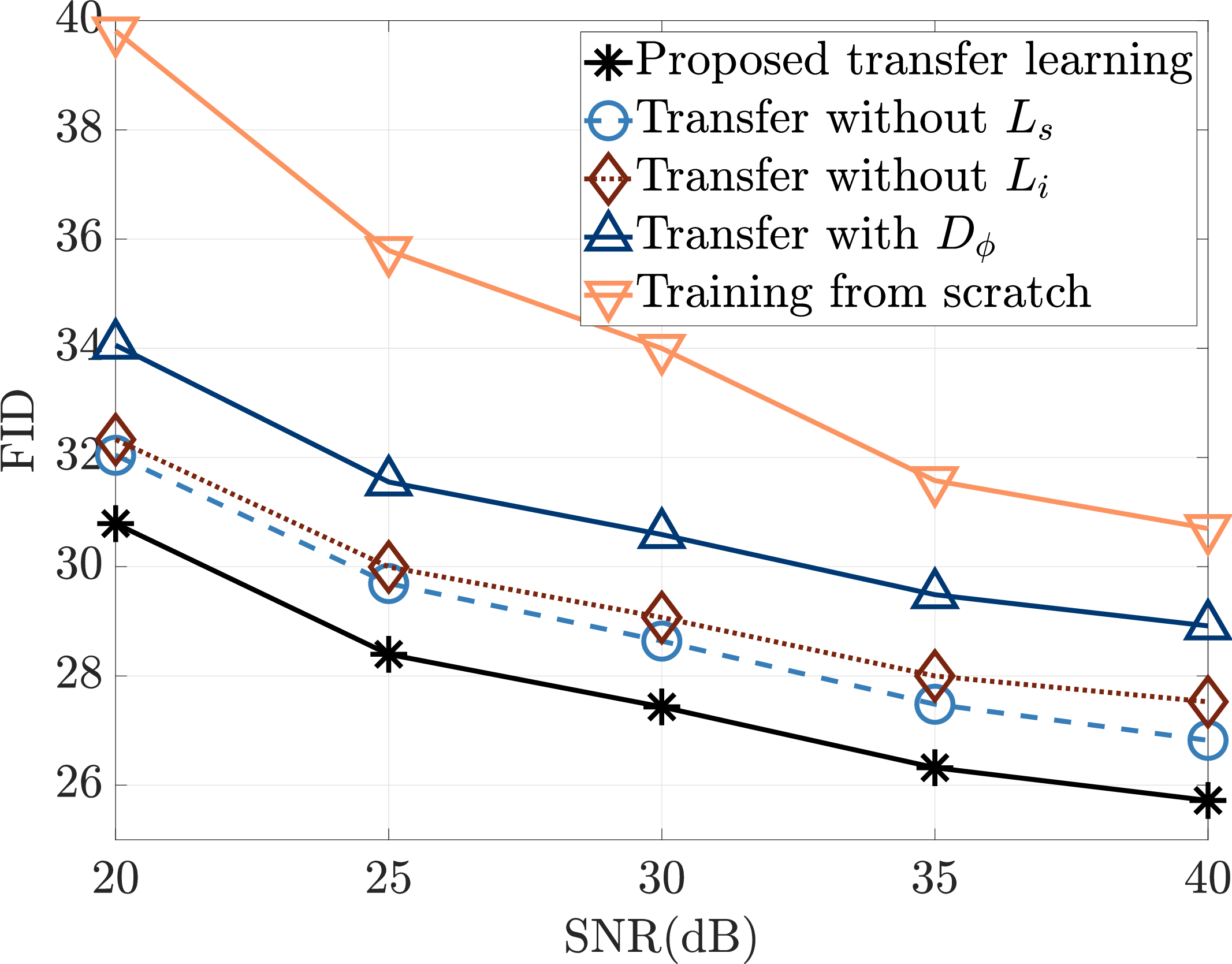}
\caption{FID performance of the proposed federated transfer multi-view synthesizing model with different image SNR.}
\label{transfer}
\end{figure}

Fig. \ref{transfer} demonstrates the performance of proposed federated transfer learning, where all models are trained with fixed epochs and communication rounds. From the figure, training from scratch performs the worst for the limited size of datasets and training time. Lack of data leads to less knowledge being learned and longer training time is required. The small size of local data also causes the model to overfit, which further degrades the performance. Then, transfer learning with the original discriminator in GAN makes the model hard to converge. As the discriminator is well-trained for the source domain, both the discriminator and generator perform domain adaptation along with GAN loss can lead to performance loss, even model collapse. We also test the proposed federated transfer learning without perceptual loss and geometry loss, both of which lead to a certain performance loss. Geometry loss weighs less but plays an important role in constructing high-quality content and preventing visual distinction. It also should be noted that training the model without freezing the camera layers will cause 3D inconsistency for the proposed VR content delivery network.

\begin{figure}[t]
\centering
\includegraphics[width = \linewidth]{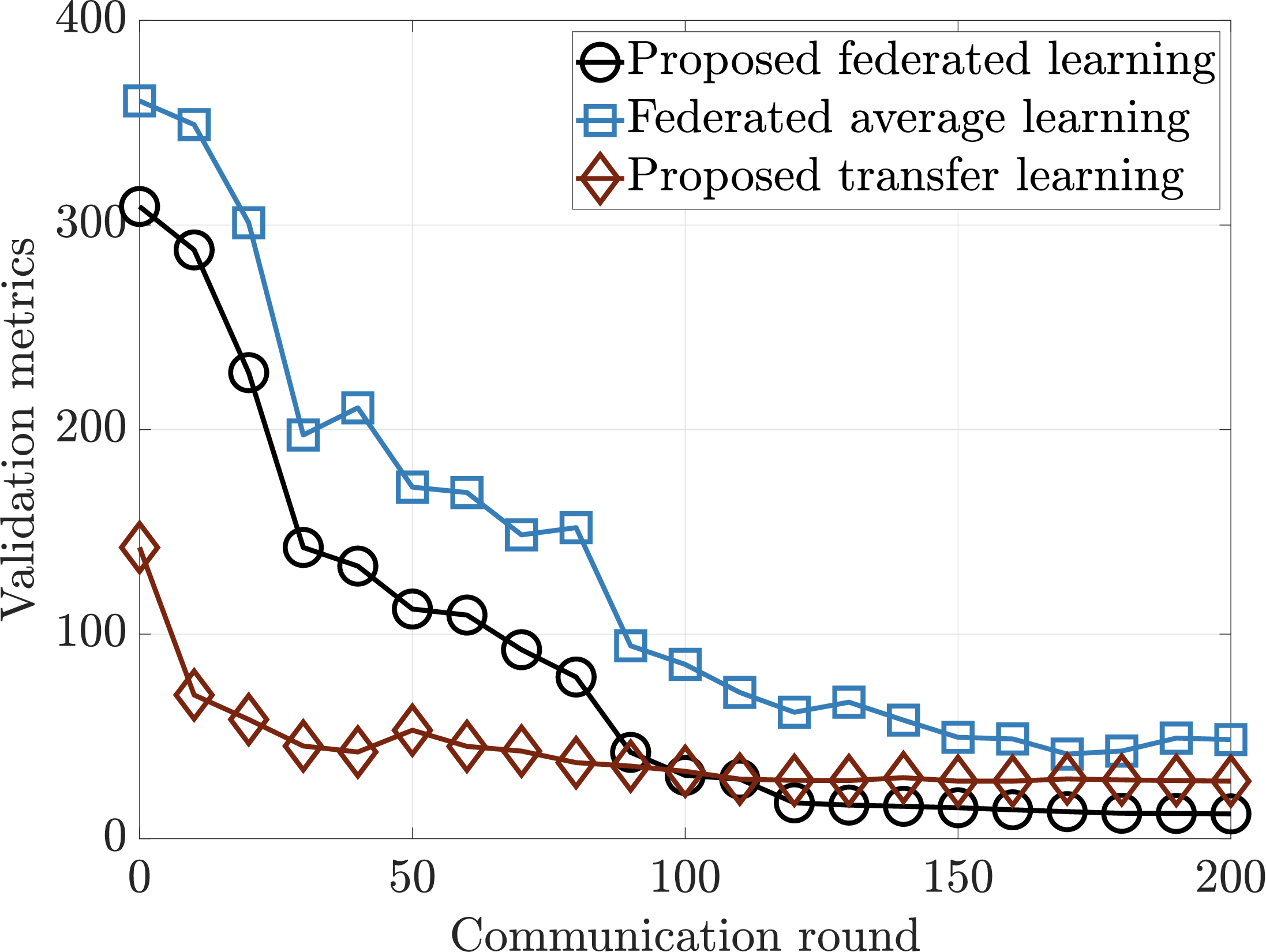}
\caption{FID performance of the proposed federated transfer multi-view synthesizing model with different image SNR.}
\label{trans}
\end{figure}

Fig. \ref{trans} shows the performance of the proposed federated learning convergence. From the figure, the proposed federated learning scheme utilizing horizontal and vertical federated learning achieves better performance than the federated average method. Both sufficient feature expression in horizontal datasets and hyper-network benefit the convergent speed. Note that the latency of each communication round for the proposed federated learning scheme is different from the federated average method. The proposed federated learning scheme only updates the reduced amount of the generator weights. Thus, the proposed federated learning scheme can support more participating devices under the same latency constraint. Also, the proposed federated transfer learning can reuse the knowledge from the pre-trained multi-view synthesizing model and achieve the best convergence speed. The proposed federated transfer learning experiences vibrate around communication 50 because no parameter is freezing. The size of the datasets limits the final performance of the proposed federated transfer learning.

\section{Conclusion}
In this paper, we proposed a federated multi-view synthesizing method to support VR services. The 3D-aware synthesizing model in the network uses only collections of single-view images to synthesize multi-view consistent images. The proposed method can serve VR users with a range of FoV by multicasting the VR content, which can improve bandwidth efficiency and latency performance for massive connections. With only single-view input given, the synthesized novel view content is not precisely the same as the actual object. The proposed model can achieve a high level of multi-view consistency. However, a definitive way to quantify the deviation from the actual content and the trade-off between multi-view consistency and accuracy based on single-view input remains an open challenge. The proposed method also reduces communication overhead and depends less on viewport prediction accuracy. The feature space and ID space of the dataset are utilized as horizontal and vertical datasets. Each client only updates part of their model based on the datasets they have. Such a method reduces the latency in the federated learning process and involves more users joining the training. To provide VR content delivery to users with insufficient datasets, we proposed a novel loss function for federated transfer learning. The proposed federated multi-view synthesizing scheme can serve multiple VR users simultaneously with the ability of novel view synthesizing.

\bibliographystyle{IEEEtran}
\bibliography{bib2021}

\begin{thebibliography}{10}
\providecommand{\url}[1]{#1}
\csname url@samestyle\endcsname
\providecommand{\newblock}{\relax}
\providecommand{\bibinfo}[2]{#2}
\providecommand{\BIBentrySTDinterwordspacing}{\spaceskip=0pt\relax}
\providecommand{\BIBentryALTinterwordstretchfactor}{4}
\providecommand{\BIBentryALTinterwordspacing}{\spaceskip=\fontdimen2\font plus
\BIBentryALTinterwordstretchfactor\fontdimen3\font minus \fontdimen4\font\relax}
\providecommand{\BIBforeignlanguage}[2]{{%
\expandafter\ifx\csname l@#1\endcsname\relax
\typeout{** WARNING: IEEEtran.bst: No hyphenation pattern has been}%
\typeout{** loaded for the language `#1'. Using the pattern for}%
\typeout{** the default language instead.}%
\else
\language=\csname l@#1\endcsname
\fi
#2}}
\providecommand{\BIBdecl}{\relax}
\BIBdecl

\bibitem{10012859}
Y.~Guo and Z.~Qin, ``{Federated learning for multi-view synthesizing in wireless virtual reality networks},'' in \emph{Proc. IEEE 96th Veh. Technol. Conf.}, 2022, pp. 1--5.

\bibitem{9430902}
P.~Lin, Q.~Song, F.~R. Yu, D.~Wang, A.~Jamalipour, and L.~Guo, ``{Wireless virtual reality in beyond 5G systems with the internet of intelligence},'' \emph{{IEEE} Wireless Commun.}, vol.~28, no.~2, pp. 70--77, 2021.

\bibitem{9711518}
A.~Moerman, J.~Van~Kerrebrouck, O.~Caytan, I.~L. de~Paula, L.~Bogaert, G.~Torfs, P.~Demeester, H.~Rogier, and S.~Lemey, ``{Beyond 5G without obstacles: mmWave-over-fiber distributed antenna systems},'' \emph{{IEEE} Commun. Mag.}, vol.~60, no.~1, pp. 27--33, 2022.

\bibitem{li2022internet}
K.~Li, Y.~Cui, W.~Li, T.~Lv, X.~Yuan, S.~Li, W.~Ni, M.~Simsek, and F.~Dressler, ``{When internet of things meets metaverse: Convergence of physical and cyber worlds},'' \emph{arXiv preprint arXiv:2208.13501}, 2022.

\bibitem{9133103}
A.~Yaqoob, T.~Bi, and G.-M. Muntean, ``{A survey on adaptive $360^{\circ}$ video streaming: solutions, challenges and opportunities},'' \emph{IEEE Commun. Surv. Tutor.}, vol.~22, no.~4, pp. 2801--2838, Jul. 2020.

\bibitem{9894271}
L.~Zhao, Y.~Cui, S.~Yang, and S.~Shamai~Shitz, ``{An optimization framework for general rate splitting for general multicast},'' \emph{{IEEE} Trans. Wireless Commun.}, vol.~22, no.~3, pp. 1573--1587, 2023.

\bibitem{10000901}
X.~Peng, Z.~Qin, D.~Huang, X.~Tao, J.~Lu, G.~Liu, and C.~Pan, ``{A robust deep learning enabled semantic communication system for text},'' in \emph{Proc. IEEE Global Commun. Conf. (GLOBECOM)}, 2022, pp. 2704--2709.

\bibitem{zhang2023semantic}
B.~Zhang, Z.~Qin, and G.~Y. Li, ``{Semantic communications with variable-length coding for extended reality},'' \emph{arXiv preprint arXiv:2302.08645}, 2023.

\bibitem{mildenhall2020nerf}
B.~Mildenhall, P.~P. Srinivasan, M.~Tancik, J.~T. Barron, R.~Ramamoorthi, and R.~Ng, ``{NeRF: representing scenes as neural radiance fields for view synthesis},'' in \emph{Proc. Eur. Conf. Comput. Vis., ECCV}, 2020, pp. 405--421.

\bibitem{cao2022scenerf}
A.-Q. Cao and R.~de~Charette, ``{SceneRF: Self-supervised monocular 3D scene reconstruction with radiance fields},'' \emph{arXiv preprint arXiv:2212.02501}, 2022.

\bibitem{cao2023communication}
X.~Cao, T.~Ba{\c{s}}ar, S.~Diggavi, Y.~C. Eldar, K.~B. Letaief, H.~V. Poor, and J.~Zhang, ``{Communication-efficient distributed learning: An overview},'' \emph{{IEEE} J. Sel. Areas Commun.}, 2023.

\bibitem{tao2023content}
Y.~Tao, Y.~Jiang, F.-C. Zheng, Z.~Wang, P.~Zhu, M.~Tao, D.~Niyato, and X.~You, ``{Content popularity prediction based on quantized federated bayesian learning in fog radio access networks},'' \emph{{IEEE} Trans. Commun.}, 2023.

\bibitem{yang2019federated}
Q.~Yang, Y.~Liu, T.~Chen, and Y.~Tong, ``{Federated machine learning: Concept and applications},'' \emph{ACM Trans. Intell. Syst. Technol.}, vol.~10, no.~2, pp. 1--19, 2019.

\bibitem{9789210}
Z.~Chen, H.~Zhu, L.~Song, D.~He, and B.~Xia, ``{Wireless multiplayer interactive virtual reality game systems with edge computing: modeling and optimization},'' \emph{{IEEE} Trans. Wireless Commun.}, pp. 1--1, 2022.

\bibitem{9565222}
X.~Liu, Y.~Deng, C.~Han, and M.~D. Renzo, ``{Learning-based prediction, rendering and transmission for interactive virtual reality in RIS-assisted terahertz networks},'' \emph{{IEEE} J. Sel. Areas Commun.}, vol.~40, no.~2, pp. 710--724, 2022.

\bibitem{gupta2022mmwave}
S.~Gupta, J.~Chakareski, and P.~Popovski, ``{mmWave networking and edge computing for scalable $360^{\circ}$ video multi-user virtual reality},'' \emph{{IEEE} Trans. Image Process.}, vol.~32, pp. 377--391, 2022.

\bibitem{van2022edge}
D.~Van~Huynh, S.~R. Khosravirad, A.~Masaracchia, O.~A. Dobre, and T.~Q. Duong, ``{Edge intelligence-based ultra-reliable and low-latency communications for digital twin-enabled metaverse},'' \emph{{IEEE} Commun. Lett.}, vol.~11, no.~8, pp. 1733--1737, 2022.

\bibitem{dang2022low}
T.~Dang, C.~Liu, and M.~Peng, ``{Low-latency mobile virtual reality content delivery for unmanned aerial vehicle-enabled wireless networks with energy constraints},'' \emph{{IEEE} Trans. Veh. Technol.}, 2022.

\bibitem{chaccour2022can}
C.~Chaccour, M.~N. Soorki, W.~Saad, M.~Bennis, and P.~Popovski, ``{Can terahertz provide high-rate reliable low-latency communications for wireless VR?}'' \emph{IEEE Internet Things J.}, vol.~9, no.~12, pp. 9712--9729, 2022.

\bibitem{10124955}
M.~Li, J.~Gao, C.~Zhou, X.~Shen, and W.~Zhuang, ``{User dynamics-aware edge caching and computing for mobile virtual reality},'' \emph{IEEE J. Sel. Top. Signal Process.}, pp. 1--13, 2023.

\bibitem{9567690}
R.~Zhang, J.~Liu, F.~Liu, T.~Huang, Q.~Tang, S.~Wang, and F.~R. Yu, ``{Buffer-aware virtual reality video streaming with personalized and private viewport prediction},'' \emph{{IEEE} J. Sel. Areas Commun.}, vol.~40, no.~2, pp. 694--709, 2022.

\bibitem{liu2023rendered}
C.~Liu, K.~Wang, H.~Zhang, X.~Li, and H.~Ji, ``{Rendered tile reuse scheme based on FoV prediction for MEC-assisted wireless VR service},'' \emph{IEEE Trans. Netw. Sci. Eng.}, 2023.

\bibitem{struye2022covrage}
J.~Struye, F.~Lemic, and J.~Famaey, ``{CoVRage: Millimeter-wave beamforming for mobile interactive virtual reality},'' \emph{{IEEE} Trans. Wireless Commun.}, 2022.

\bibitem{9798771}
P.~Yang, T.~Q.~S. Quek, J.~Chen, C.~You, and X.~Cao, ``{Feeling of presence maximization: mmWave-enabled virtual reality meets deep reinforcement learning},'' \emph{{IEEE} Trans. Wireless Commun.}, pp. 1--1, 2022.

\bibitem{wang20224k}
Z.~Wang, L.~Li, Z.~Shen, L.~Shen, and L.~Bo, ``{4K-NeRF: High fidelity neural radiance fields at ultra high resolutions},'' \emph{arXiv preprint arXiv:2212.04701}, 2022.

\bibitem{li2022compressing}
L.~Li, Z.~Shen, Z.~Wang, L.~Shen, and L.~Bo, ``{Compressing volumetric radiance fields to 1 MB},'' \emph{arXiv preprint arXiv:2211.16386}, 2022.

\bibitem{gu2021stylenerf}
J.~Gu, L.~Liu, P.~Wang, and C.~Theobalt, ``{Stylenerf: a style-based 3D-aware generator for high-resolution image synthesis},'' \emph{arXiv preprint arXiv:2110.08985}, 2021.

\bibitem{pan2020gan2shape}
X.~Pan, B.~Dai, Z.~Liu, C.~C. Loy, and P.~Luo, ``{Do 2D GANs know 3D shape? Unsupervised 3D shape reconstruction from 2D image GANs},'' in \emph{Proc. Int. Conf. Learn. Representations, ICLR}, 2021.

\bibitem{kwak2022injecting}
J.-g. Kwak, Y.~Li, D.~Yoon, D.~Kim, D.~Han, and H.~Ko, ``{Injecting 3D perception of controllable NeRF-GAN into stylegan for editable portrait image synthesis},'' in \emph{Proc. Eur. Conf. Comput. Vis., ECCV}.\hskip 1em plus 0.5em minus 0.4em\relax Springer, 2022, pp. 236--253.

\bibitem{cai2022pix2nerf}
S.~Cai, A.~Obukhov, D.~Dai, and L.~Van~Gool, ``{Pix2nerf: Unsupervised conditional $\pi$-GAN for single image to neural radiance fields translation},'' in \emph{Proc. IEEE Conf. Comput. Vis. Pattern Recognit., CVPR}, 2022, pp. 3981--3990.

\bibitem{acar2020federated}
D.~A.~E. Acar, Y.~Zhao, R.~Matas, M.~Mattina, P.~Whatmough, and V.~Saligrama, ``{Federated learning based on dynamic regularization},'' in \emph{Proc. Int. Conf. Learn. Representations, ICLR}, 2020.

\bibitem{thapa2022splitfed}
C.~Thapa, P.~C.~M. Arachchige, S.~Camtepe, and L.~Sun, ``{Splitfed: When federated learning meets split learning},'' in \emph{Proc. AAAI Conf. Artif. Intell.}, vol.~36, no.~8, 2022, pp. 8485--8493.

\bibitem{zhao2022generative}
X.~Zhao, F.~Ma, D.~G{\"u}era, Z.~Ren, A.~G. Schwing, and A.~Colburn, ``{Generative multiplane images: Making a 2D GAN 3D-aware},'' \emph{arXiv preprint arXiv:2207.10642}, 2022.

\bibitem{skorokhodov2022epigraf}
I.~Skorokhodov, S.~Tulyakov, Y.~Wang, and P.~Wonka, ``{EpiGRAF: Rethinking training of 3D GANs},'' \emph{arXiv preprint arXiv:2206.10535}, 2022.

\bibitem{goodfellow2014generative}
I.~Goodfellow, J.~Pouget-Abadie, M.~Mirza, B.~Xu, D.~Warde-Farley, S.~Ozair, A.~Courville, and Y.~Bengio, ``{Generative adversarial nets},'' in \emph{Proc. Adv. Neural Inf. Process. Syst. (NIPS)}, vol.~27, 2014.

\bibitem{chan2022efficient}
E.~R. Chan, C.~Z. Lin, M.~A. Chan, K.~Nagano, B.~Pan, S.~De~Mello, O.~Gallo, L.~J. Guibas, J.~Tremblay, S.~Khamis \emph{et~al.}, ``{Efficient geometry-aware 3D generative adversarial networks},'' in \emph{Proc. IEEE Conf. Comput. Vis. Pattern Recognit., CVPR}, 2022, pp. 16\,123--16\,133.

\bibitem{chan2021pi}
E.~R. Chan, M.~Monteiro, P.~Kellnhofer, J.~Wu, and G.~Wetzstein, ``{Pi-GAN: Periodic implicit generative adversarial networks for 3D-aware image synthesis},'' in \emph{Proc. IEEE Conf. Comput. Vis. Pattern Recognit., CVPR}, 2021, pp. 5799--5809.

\bibitem{kolouri2018sliced}
S.~Kolouri, G.~K. Rohde, and H.~Hoffmann, ``{Sliced wasserstein distance for learning gaussian mixture models},'' in \emph{Proc. IEEE Conf. Comput. Vis. Pattern Recognit., CVPR}, 2018, pp. 3427--3436.

\bibitem{zhang2022generalized}
Z.~Zhang, Y.~Liu, C.~Han, T.~Guo, T.~Yao, and T.~Mei, ``{Generalized one-shot domain adaption of generative adversarial networks},'' \emph{arXiv preprint arXiv:2209.03665}, 2022.

\bibitem{karras2019style}
T.~Karras, S.~Laine, and T.~Aila, ``{A style-based generator architecture for generative adversarial networks},'' in \emph{Proc. IEEE Conf. Comput. Vis. Pattern Recognit., CVPR}, 2019, pp. 4401--4410.

\bibitem{karras2020training}
T.~Karras, M.~Aittala, J.~Hellsten, S.~Laine, J.~Lehtinen, and T.~Aila, ``{Training generative adversarial networks with limited data},'' \emph{Adv. Neural Inf. Process. Syst.}, vol.~33, pp. 12\,104--12\,114, 2020.

\bibitem{heusel2017gans}
M.~Heusel, H.~Ramsauer, T.~Unterthiner, B.~Nessler, and S.~Hochreiter, ``{GANs trained by a two time-scale update rule converge to a local nash equilibrium},'' \emph{Proc. NeurIPS}, vol.~30, 2017.

\bibitem{binkowski2018demystifying}
M.~Bi{\'n}kowski, D.~J. Sutherland, M.~Arbel, and A.~Gretton, ``{Demystifying mmd GANs},'' \emph{arXiv preprint arXiv:1801.01401}, 2018.

\bibitem{perfecto2020taming}
C.~Perfecto, M.~S. Elbamby, J.~Del~Ser, and M.~Bennis, ``{Taming the latency in multi-user VR 360°: A QoE-aware deep learning-aided multicast framework},'' \emph{{IEEE} Trans. Commun.}, vol.~68, no.~4, pp. 2491--2508, 2020.

\end{thebibliography}

\end{document}